\documentclass{aa}
\usepackage{txfonts}
\usepackage{graphicx}
\usepackage{aalongtable}
\begin{document}
\title{CNO in evolved intermediate mass stars
\thanks{Observations collected at ESO,La Silla, Chile} }
\author{
R. Smiljanic\inst{1}
\and
B. Barbuy\inst{1}
\and
J. R. De Medeiros\inst{2}
\and
A. Maeder\inst{3}
}
\offprints{B. Barbuy}
\institute{
Universidade de S\~ao Paulo, IAG, Rua do Mat\~ao 1226,
Cidade Universit\'aria, 05508-900, S\~ao Paulo, SP, Brazil\\
 e-mail: rodolfo@astro.iag.usp.br, barbuy@astro.iag.usp.br
\and
Universidade Federal do Rio Grande do Norte, Departamento de
F\'isica, Campus Universit\'ario, 59072-970, Natal, RN, Brazil\\
e-mail: renan@dfte.ufrn.br
\and
Geneva Observatory, 1290 Sauverny, Switzerland \\
e-mail: Andre.Maeder@obs.unige.ch
}
\date{Received / Accepted}
\abstract{
In order to investigate the possible influence of rotation on the efficiency of the first dredge-up we determined atmospheric parameters, masses, and abundances of carbon, nitrogen, and oxygen in a sample of evolved intermediate mass stars. We used high resolution spectra and conducted a model atmosphere analysis. The abundances were calculated through spectral synthesis and compared to the predictions of rotating and non-rotating evolutionary models. Almost all those objects in our sample where carbon and nitrogen abundances could be determined show signs of internal mixing. The stars, however, seem to be mixed to different extents. Among the mixed stars we identify five in our sample with abundances in agreement with the non-rotating models, four stars that seem to be mixed beyond that, and one star that seems to be slightly less mixed than predicted for the first dredge-up. There are also five stars that seem to be slightly more mixed than expected, but their abundances are in marginal agreement with both rotating and non-rotating models. Such differences in the extent of the mixing are not predicted by the standard models and imply the action of other mixing mechanisms than solely the convective dredge-up. We also identified for the first time an important correlation between the [N/C] ratio and the stellar mass.
}
\titlerunning{CNO in evolved intermediate mass stars}
\authorrunning{Smiljanic et al.}
\maketitle


\section{Introduction} 

The evolution of a star is usually treated as a function only of initial mass and chemical composition. Issues such as rotation and magnetic fields are considered as playing a minor role. However, in the past few years discrepancies have been found between model predictions and the observations of abundances in intermediate mass stars (5-20M$_{\odot}$), which could be due to the earlier neglect of rotation.
\par Intermediate mass stars burn hydrogen during the main sequence (MS) via the CNO cycle. One of the main outcomes of the CNO cycle is the conversion of almost all central C$^{12}$ into N$^{14}$. When the central hydrogen is exhausted, the star expands its outer layers and cools off, evolving rapidly to the red giant branch (RGB). Before it starts to burn He, the star experiences the so-called first dredge-up, the development of a deep convective layer that brings nuclear processed material to the surface. The photospheric abundances of carbon and nitrogen are then altered (C is reduced and N is increased). 
\par During the RGB, depending on its mass, the star can experience a blue-loop; in the models by Schaller et al. (\cite{Sc92}) the evolutionary tracks for stars between 5 and 12 M$_{\odot}$ have blue loops. The star evolves from the RGB to the blue-giant region, due to a temporary increase in the effective temperature, and then back to the RGB. The occurrence and extent of the loop are dependent on the mass and on other uncertain factors, such as the treatment of convection. More details on the blue loops can be found in Xu \& Li (\cite{XL04}) and references therein. Since the first crossing of the HR diagram, from the MS to the RGB, happens on a short time scale of a few million years, most of the stars observed in the blue giant region must be evolving through a blue loop. 
\par In spite of this relatively simple description, the observations reveal a more complex scenario. Although an important mixing episode is not supposed to happen before the first dredge-up there is evidence of He (Lyubimkov \cite{Ly98}) and N (Gies \& Lambert \cite{GL92} and Lennon et al.\ \cite{L96}) overabundances in O and B stars. In addition, boron seems to be highly depleted in main sequence B stars (Venn et al. \cite{Venn02}). These are probable indications of mixing during the MS. In particular, Fliegner et al.\ (\cite{F96}) show that it is possible to qualitatively reproduce the boron behavior by including rotation effects in theoretical evolutionary calculations.
\par On leaving the MS, as well as during the blue loop, the stars will eventually shine as A type stars. Venn (\cite{V95a,V95b}) analyzed a sample of Galactic A-type supergiants and found stars with unchanged composition, as well as stars with slightly modified composition. These stars are probably crossing the HR diagram for the first time and the latter group might be reflecting a mixing episode that occurred during the MS. Venn (\cite{V99}) analyzed A-type supergiants from the SMC where some stars showed signs of the first dredge-up and others were non-mixed. The nitrogen abundance in the post first dredge-up stars shows a spread that is not predicted by the models. This spread might probably be due to different rotations inducing different mixing efficiencies.
\par There are only a few abundance determinations in yellow (F-G) and red (K-M) supergiants. Sodium overabundances were detected (Boyarchuk \& Lyubimkov \cite{BL83}) and they appear to be correlated with mass (Sasselov \cite{S86}). The Na overabundances are probably related to the operation of the Ne-Na cycle during hydrogen burning. El Eid \& Champagne (\cite{EC95}) have investigated the Ne-Na cycle theoretically in A-F supergiants and their results agree well with observations.
\par Luck \& Lambert (\cite{LL85}) determined C, N, and O abundances in a sample of 2 variables and 4 non-variables F supergiants. They found a higher N/C ratio than expected (more N and less C), which is an indication of a more efficient mixing in these stars. Barbuy et al.\ (\cite{B96}) have determined C, N, and O abundances in a sample of 9 low-rotator F supergiants. They found stars with non-modified abundances and stars with abundances that are only slightly changed.
\par There are also some determinations of C and N abundances in Cepheid stars (Andrievsky et al.\ \cite{A96}, \cite{A02}, \cite{A04}, Luck et al.\ \cite{L03}, Kovtyukh et al.\ \cite{K96},  Usenko et al.\ \cite{U01a,U01b}). In general there are stars without changes in their abundances, which are probably crossing the instability strip for the first time, stars with [N/C] near the expected value for the first dredge-up, and stars with abundances changed beyond what is expected. The last group of stars has probably passed through more efficient mixing processes. It is worth noting that Kovtyukh et al.\ (\cite{K96}) have found two stars overabundant in Na but with normal abundances of C and N.
\par Consequently, the scenario suggested by the observations is much more complex than predicted by the standard models. Two characteristics in particular must be stressed. The first is the indication of mixing processes during the MS. The second is the indication of a more efficient mixing than the expected solely due to the first dredge-up. Neither are predicted by the standard non-rotating models. The inclusion of rotation in the models seems to be able to reproduce these behaviors at least qualitatively.
\par Much effort has been made in the last years towards a better physical treatment of rotation and its effects, as reviewed by Maeder \& Meynet (\cite{MaM00}) and references therein. Effects induced by rotation, such as meridional circulation (Maeder \& Zahn \cite{MZ98}) and especially shear turbulence (Maeder \cite{M97}; Mathis et al.\ \cite{M04}; Mathis \& Zahn \cite{MZ04}) act in transporting and mixing the chemical elements. Thus the required additional mixing mechanism seems to appear naturally when rotation effects are taken into account. Even a mixing event during the MS should happen if the rotation is sufficiently high. Although this seems promising, much work still needs to be done. More observational results are very needed in order to better constrain the models.
\par In this work we derive C, N, and O abundances in a sample of cool giants and supergiants. The observations are described in  Sect. 2, the stellar parameters are described in Sect. 3, and the abundances of CNO are described in Sect. 4. The results are discussed in Sect. 5 and conclusions drawn in Sect. 6.

\section {Observations} 

High resolution CCD spectra were obtained for a sample of 19 cool luminous stars using the FEROS spectrograph at the ESO 1.52m telescope at La Silla (Chile). FEROS is a fiber-fed echelle spectrograph that provides a full wavelength coverage of {$\lambda\lambda$} 3500-9200 \AA \,over 39 orders at a resolving power of R=48,000 (Kaufer et al.\ \cite{K00}). The detector used was an EEV CCD chip with 2048x4096 pixels and with a pixel size of 15$\mu$m. The program stars were observed during four observational runs in 2000 and 2001, as given in Table \ref{table1}. All spectra were reduced using the FEROS pipeline software. 
\par An estimation of the average signal to noise ratio for each spectrum is given in Table \ref{table1}. This table also lists the spectral type, visual magnitude, galactic coordinates, parallax, heliocentric radial velocity, and rotational velocity of the program stars. The rotational velocities were mainly taken from De Medeiros et al.\,(\cite{M02}) with the exception of \object{HD 80404} from Royer et al.\ (\cite{Ro02}) and \object{HD 38713} and \object{HD 45348} from De Medeiros (2005, private communication). The radial velocities were determined using IRAF. The other stellar data in Table \ref{table1} were taken from Simbad\footnote{This research made use of the Simbad database operated at the CDS, Strasbourg, France.}.
\par The equivalent widths of FeI and FeII lines were determined by fitting Gaussian 
profiles to the lines with IRAF, and in the analysis, lines with equivalent widths larger than 150m{\AA} were not used. For the three hottest stars, \object{HD 36673}, HD 45348, and HD 80404, no lines larger than 100m{\AA} were used.
\begin{table*}
\caption{Sample stars: HD number, spectral type, visual magnitude, galactic coordinates, parallax, rotational velocity, heliocentric radial velocity of the stars, signal to noise ratio and, date of observation.}
\label{table1}
\centering
\begin{tabular}{c c c c c c c c c c}
\noalign{\smallskip}
\hline\hline
\noalign{\smallskip}
HD & ST & $V$ & $b$ & $l$ & $\pi$ & v $sin i$ & V$^{hel.}_{r}$ & $S/N$ & Date of\\
   &    &     &     &     &       &   km s$^{-1}$    & km s$^{-1}$ &       & obs.\\
\hline
1219 & K1IV & 8.91 & $-$64.6$^\circ$ & 315.7$^\circ$ & 5.30 & -- & $-$19.6 & 250 & 10.19.2000 \\
36673 & F0Ib & 2.60 & $-$25.1$^\circ$ & 220.9$^\circ$ & 2.54 & 10.0 & 2.4 & 450 & 02.17.2000 \\
38713 & G8III & 6.17 & $-$21.4$^\circ$ & 220.8$^\circ$ & 4.91 & 2.3 & 6.3 & 300 & 01.15.2001 \\
44362 & G2Ib & 7.04 & $-$25.6$^\circ$ & 258.5$^\circ$ & 1.24 &  8.8 & 14.9 & 320 & 01.15.2000 \\
45348 & F0II & $-$0.72 & $-$25.3$^\circ$ & 261.2$^\circ$ & 10.43 & 8.0 & 20.6 & 420 & 01.15.2000 \\
49068 & K0-1III & 7.43 & $-$10.6$^\circ$ & 231.1$^\circ$ & 2.18 & $<$1.0 & 23.8 & 400 & 01.15.2000 \\
49396 & G6Iab & 6.55 & $-$22.0$^\circ$ & 261.6$^\circ$ & 1.49 & 8.6 & 29.4 & 290 & 01.15.2000 \\
51043 & G5Ib-II & 6.56 & $-$21.5$^\circ$ & 263.9$^\circ$ & 2.36 & 3.5 & 14.3 & 210 & 01.15.2000 \\
66190 & K1Ib-II & 6.61 & $-$8.1$^\circ$ & 260.4$^\circ$ & 0.60 & 4.0 & 27.4 & 360 & 01.15.2000 \\
71181 & G6Ib-II & 7.62 & $-$4.4$^\circ$ & 262.4$^\circ$ & $-$1.80 & 2.1 & 13.5 & 550 & 02.14.2000\\
76860 & K3Ib & 7.14 & $-$2.7$^\circ$ & 269.4$^\circ$ & 0.43 & 2.0 & 4.5 & 460 & 01.16.2001 \\
80404 & A8Ib & 2.25 & $-$7.0$^\circ$ & 278.5$^\circ$ & 4.71 & 10.0 & 10.9 & 330 & 01.16.2001 \\
90289 & K4III & 6.34 & $-$0.5$^\circ$ & 284.4$^\circ$ & 4.11 & $<$1.0 & $-$16.1 & 250 & 01.14.2001 \\
102839 & G5Ib & 4.98 & $-$8.0$^\circ$ & 297.7$^\circ$ & 2.24 & 7.6 & 14.2 & 240 & 02.14.2000 \\
114792 & F5-F6Ib & 6.85 & +0.1$^\circ$ & 305.4$^\circ$ & 0.36 & 7.5 & $-$17.5 & 410 & 02.14.2000 \\
159633 & G2Ib & 6.27 & $-$3.4$^\circ$ & 351.3$^\circ$ & 1.15 & 9.1 & 11.6 & 310 & 10.04.2001 \\
192876 & G3Ib & 4.25 & $-$24.7$^\circ$ & 31.1$^\circ$ & 4.75 & 7.3 & $-$27.4 & 350 & 10.20.2000 \\
204867 & G0Ib & 2.91 & $-$37.9$^\circ$ & 48.0$^\circ$ & 5.33 & 9.5 & 6.1 & 360 & 10.20.2000 \\
225212 & K3Iab & 4.95 & $-$70.0$^\circ$ & 87.1$^\circ$ & 2.03 & 5.8 & $-$42.0 & 320 & 10.19.2000 \\
\noalign{\smallskip}
\hline
\end{tabular}
\end{table*}


\section{Stellar parameters}

\subsection{Physical data}

Oscillator strengths for FeI lines were prefered from the NIST database (Martin et al. \cite{NIST}) complemented
by the list used by Barbuy et al.\,(\cite{B96}). For the FeII lines the oscillator 
strengths were mainly taken from Mel\'endez\,\&\,Barbuy\,(\cite{MB05}) complemented by Barbuy 
et al.\,(\cite{B96}) and Kovtyukh\,\&\,Andrievsky\,(\cite{KA99}). The equivalent widths and oscillator strengths are listed in the Appendix (Tables \ref{LE}, \ref{LE2}, \ref{LE3}, and \ref{LE4}).
\par In this analysis we used grids of model atmospheres generated by the ATLAS9 code\,(Kur\'ucz \cite{K94}) for stars hotter than 4750K, whereas grids of model atmospheres by the NMARCS code\,(Plez et al. \cite{P92}) were adopted for the cooler stars. The ATLAS9  models assume local thermodynamic equilibrium (LTE), plane-parallel geometry, and hydrostatic equilibrium. The NMARCS models are spherically symmetric and assume LTE and hydrostatic equilibrium. The NMARCS models are a more suitable choice for the cooler stars for its better description of the opacity sources. Whenever necessary, codes for interpolating among the grids were adopted.

\subsection{Effective temperature}

For each star we derived the effective temperature using four different approaches: the excitation equilibrium of FeI lines, the excitation equilibrium of FeII lines, photometric calibrations, and fitting the H$\alpha$ line wings. Using each temperature estimate, we calculated the complete set of atmospheric parameters, namely surface gravity (log g), microturbulence velocity ($\xi$) and, metallicity ([Fe/H]). After a comparison we chose the more reliable set of parameters as described below.

\subsubsection{T$_{\rm{eff}}$ from FeI}

In the derivation of the T$_{\rm{eff}}$ from the FeI lines, all the parameters were calculated  simultaneously. In this method the T$_{\rm{eff}}$ was found by requiring a null correlation of the iron abundance as given by the FeI lines with the excitation potential (the excitation equilibrium). The surface gravity was found by requiring both FeI and FeII lines to have the same mean abundance (the ionization equilibrium). The microturbulence velocity was found by requiring the iron abundance (from FeI lines) to have a null correlation with the equivalent widths. By fulfilling these conditions we also determined the iron abundance, [Fe/H].
\par Usually one adopts FeI for this kind of analysis because it is a chemical species with a high number of lines available in the spectrum. However, there is evidence that it must be adopted with care in some cases. Lyubimkov \& Boyarchuk (\cite{LB83}) argue that in F supergiants FeI might be overionized due to non-LTE effects; FeII, on the other hand, should not be affected.
\par When computing the parameters with the FeI lines, we noticed a tendency for the hotter stars to have higher metallicities (around [Fe/H] = +0.3 dex). In particular, the three stars with earlier spectral type, HD 36673 (F0), HD 45348 (F0), and HD 80404 (A8) showed metallicities larger than [Fe/H] = +0.50 dex. These results do not seem to be reasonable and are probably spurious due to NLTE effects. Since FeII is probably not affected by NLTE (Lyubimkov \& Boyarchuk \cite{LB83}), its use to constrain the parameters should produce more reliable results (Kovtyukh \& Andrievsky \cite{KA99}).

\subsubsection{T$_{\rm{eff}}$ from FeII}

The parameters were recalculated through the same steps described using FeI. The main obstacle in using FeII lines is their reduced number. The number of lines used in this work varies from 6 to 18 per star. A small set of lines containing one or two unreliable lines, affected by blends or by uncertain $gfs$, can generate unreliable correlations, thus leading to uncertain parameters. That was probably the case for the stars \object{HD 66190}, \object{HD 76860}, and \object{HD 102839}, for which we found metallicities larger than [Fe/H] = +0.60 dex. According to the FeI parameters, these are cool stars where there are more blends.
\par However there are two points worth noting. First, the parameters for the three hotter stars, as derived from the FeII lines, show a good agreement with the parameters in the literature. Second, in general the metallicities obtained in this way are smaller than the ones obtained with the FeI lines. This is mainly due to a larger $\xi$ and a slightly lower T$_{\rm{eff}}$ obtained from the FeII lines.

\subsubsection{T$_{\rm{eff}}$ from H$\alpha$}

The wings of hydrogen lines are good temperature indicators considering they are independent of log g and $\xi$ for a large range of temperatures. We estimated the T$_{\rm{eff}}$ by fitting the H$\alpha$ wings with synthetic spectra. 
\par
\begin{figure}
\resizebox{\hsize}{!}{\includegraphics{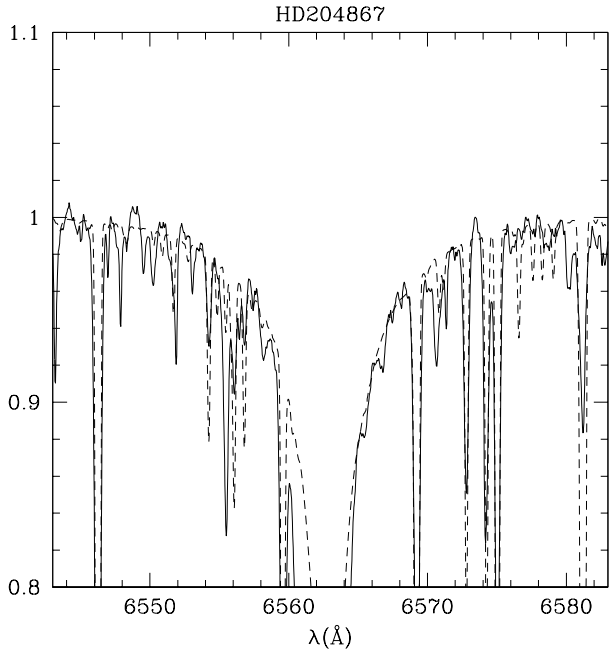}}
\caption{Fit to H$\alpha$ wings in \object{HD 204867}, the dashed line is the synthetic spectrum, and the continuous line is the observed one.}
\label{hafit}
\end{figure}

 In this method synthetic H$\alpha$ profiles are calculated for a variety of temperatures until a best fit to the observed profile is found. The synthetic spectra were calculated by programs described in Barbuy et al.\ (\cite{B03}), and the hydrogen line profile was calculated with an improved version of the code presented by Praderie (\cite{P67}). The programs calculate the H$\alpha$ profile and the lines that contaminate its wings. An example of the fit is given in Fig. \ref{hafit}.
\par For cooler stars than 4900K, the H$\alpha$ line has no pronounced wings to be fitted, hence the temperature could be determined in this way only for the hotter ones. At gravities below $2.5$ dex and higher temperatures than 7000K, the H$\alpha$ line becomes sensitive to both $T_{\rm{eff}}$ and log g. Three of our stars, HD 45348, HD 36673, and HD 80404, are in these conditions, so an independent estimate for the log g is then required. For these stars we chose the best fit with the log g as close as possible to the one derived when adopting the T$_{\rm{eff}}$ and $\xi$ from the FeII lines as discussed above.

\subsubsection{T$_{\rm{eff}}$ from photometry}

\par We also calculated a photometric estimate of the effective temperature using the $(V-K)$ calibrations of McWilliam (\cite{MW91}), Van Belle et al.\ (\cite{VB99}), and 
Houdashelt et al.\ (\cite{H00}). A mean value was calculated by adopting no weight difference. The calibration of McWilliam\ (\cite{MW91}) is valid for F supergiants but the calibrations of Van Belle et al.\ (\cite{VB99}) and Houdashelt et al.\ (\cite{H00}) are appropriate for giant stars. In spite of that we noticed a very good agreement between the temperatures derived from both calibrations as can be seen in Table \ref{table2}. 
\par Visual magnitudes were obtained from Simbad and $K_s$ magnitudes from 2MASS  (\cite{2mass}). The $K_s$ magnitudes were transformed into Johnson $K$ magnitudes by means of the relations from Alonso et al.\ (\cite{A98}). Interstellar visual extinction, A$_{V}$, was calculated using the relations from Chen et al.\ (\cite{Chen98}) for stars with $\left\vert b \right\vert < 10^\circ$ and $d < 1kpc$, and the relations from  Hakkila et al.\ (\cite{HK97}) for the other ones. In both cases the visual extinction is calculated given the galactic coordinates and distance.  The distances were calculated using the parallaxes from the Hipparcos catalogue (\cite{hip}) or from the Tycho catalogue (\cite{hip}) whenever measurements from Hipparcos were not available. Visual extinctions were transformed into color excesses adopting A$_{V}/E(B-V) = 3.10$ as the ratio of total to selective absorption. In order to deredden the color $(V-K)$ we used the expression from Rieke \& Lebofsky (\cite{RL85}), $E(V-K) = 2.744E(B-V)$. The extinction, $A_V$ and the dereddened color, $(V-K)_0$ are listed in Table \ref{table2}.
\par For one star in particular, \object{HD 114792}, the color excess was calculated using a different approach. The Hipparcos parallax indicates a distance of $d=2.78kpc$ and, since it is located in the Galactic Plane, the formula from Hakkila et al.\ predicts a rather high extinction, 
A$_V = 4.02 mag$. By using this extinction one obtains a high effective temperature, 
$T = 11000K$. However, by visual inspection, its spectrum is not of a hot star. Thus we 
concluded that the distance estimate is not adequate and a distance-independent color excess 
should be a better approach. We then adopted the $E(B-V)$ calibration of Str\"omgren photometry from Arellano-Ferro \& Parrao (\cite{AFP90}).
\begin{table*}
\caption{Interstellar extinction, dereddened color $(V-K)_0$, photometric estimates of the
effective temperature as calculated from the indicated calibrations and mean values.}
\label{table2}
\centering
\begin{tabular}{c c c c c c c}
\noalign{\smallskip}
\hline\hline
\noalign{\smallskip}
HD & A$_V$ & $(V-K)_0$ & $T_{\rm{eff}}\,(V-K)$ & $T_{\rm{eff}}\,(V-K)$ &  $T_{\rm{eff}}\,(V-K)$ & Final $T_{\rm{eff}}$ \\
   &       &           & McWilliam(1991)  &  vanBelle(1999)  & Houdashelt(2000) &  \\
\hline
1219   & 0.03 & 2.43 & -- & 4696 & 4685 & 4691 \\
36673  & 0.08 & 0.73 & 6971 & -- & 6975 & 6973 \\
38713  & 0.07 & 1.86 & -- & 5162 & 5295 & 5229 \\
44362  & 0.09 & 1.77 & 5069 & 5251 & 5411 & 5244 \\
45348  & 0.08 & 0.47 & 7446 & -- & 7448 & 7447 \\
49068  & 0.11 & 2.68 & -- & 4525 & 4468 & 4497 \\
49396  & 0.17 & 1.93 & -- & 5101 & 5214 & 5157 \\
51043  & 0.17 & 2.21 & -- & 4864 & 4903 & 4884 \\
66190  & 0.50 & 2.23 & -- & 4851 & 4886 & 4868 \\
71181  & 0.24 & 2.44 & -- & 4691 & 4678 & 4685 \\
76860  & 1.49 & 2.63 & -- & 4563 & 4515 & 4539 \\
80404  & 0.07 & 0.62 & 7173 & -- & 7172 & 7173 \\
90289  & 0.10 & 3.38 & -- & 4136 & 4029 & 4082 \\
102839 & 0.25 & 2.74 & -- & 4489 & 4423 & 4456 \\
114792 & 1.02 & 1.44 & 5662 & -- & -- & 5662 \\
159633 & 1.17 & 1.50 & 5562 & -- & 5766 & 5664 \\
192876 & 0.25 & 2.06 & -- & 4984 & 5061 & 5023 \\
204867 & 0.11 & 1.56 & 5446 & 5456 & 5679 & 5527 \\
225212 & 0.08 & 3.45 & -- & 4105 & 3999 & 4052 \\
\noalign{\smallskip}
\hline
\end{tabular}
\end{table*}

\subsection{Surface gravity}

Each temperature estimate was used to calculate a set of atmospheric parameters. In all cases the surface gravity was calculated by requiring the ionization equilibrium of FeI and FeII. The only exceptions were made when using the T$_{\rm{eff}}$ from H$\alpha$ for the stars HD 36673, HD 45348, and HD 80404. In these cases we tried to keep the gravity as close as possible to the value derived when T$_{\rm{eff}}$ and $\xi$ were determined from the FeII lines. However, the value usually needed to be adjusted during the fit itself. With a fixed log g the wings of the line do not grow indefinitely with increasing T$_{\rm{eff}}$. There is a maximum, given that further increasing of the T$_{\rm{eff}}$ will diminish the intensity of the wings. Thus, it was not always possible to keep the exact value of log g, so some adjustments were necessary.

\subsection{Microturbulence}

We determined the microturbulence from both FeI and FeII lines by requiring the abundances to have a null correlation with the equivalent width. Thus, up to six different sets of parameters were calculated for each star: i) T$_{\rm{eff}}$ and $\xi$ from FeI lines, ii) T$_{\rm{eff}}$ and $\xi$ from FeII lines, iii) T$_{\rm{eff}}$ from photometric calibrations and $\xi$ from FeI lines, iv) T$_{\rm{eff}}$ from photometric calibrations and $\xi$ from FeII lines, v) T$_{\rm{eff}}$ from H$\alpha$ and $\xi$ from FeI lines, and vi) T$_{\rm{eff}}$ from H$\alpha$ and $\xi$ from FeII lines. When comparing the parameters with the same T$_{\rm{eff}}$ but different $\xi$, we noticed that the $\xi$ from the FeII lines is usually larger, hence the metallicity is usually smaller.

\subsection{Adopted parameters}

Final atmospheric parameters were chosen among the six different estimates described above. Among all the methods used to estimate the T$_{\rm{eff}}$, the H$\alpha$ fitting is the most reliable. It has problems with neither NLTE nor reddening corrections, despite some uncertainty coming from the model atmosphere (Castilho et al. \cite{C00}). Thus we adopted the T$_{\rm{eff}}$ from H$\alpha$ for all those stars where it could be determined. Moreover, since the FeI lines are not reliable for determining $\xi$ in the hot stars, we adopted the $\xi$ as given by the FeII lines for all those stars for which T$_{\rm{eff}}$ was derived from H$\alpha$.
\par For the cool stars, where H$\alpha$ fitting was not possible, we adopted the parameters as given by the FeI lines. In these stars NLTE is not expected to be significant, and the FeII proved not to inspire much confidence, probably because of an increased number of blends.
\par There was only one exception. Even though \object{HD 1219} is a cool star, we adopted the parameters as given by the FeII lines. This choice was made because the parameters derived from the FeI lines proved unreliable. The metallicity as derived from FeI lines was [Fe/H] $>$ +0.50 dex. Table \ref{tab:adopted} lists the adopted parameters and the method of calculation.

\begin{table*}
\caption{Adopted atmospheric parameters.}
\label{tab:adopted}
\centering
\begin{tabular}{c c c c r r c}
\noalign{\smallskip}
\hline\hline
\noalign{\smallskip}
HD & $T_{\rm{eff}}$ & log g & $\xi$ & [FeI/H]$\pm\sigma$\,(\#) & [FeII/H]$\pm\sigma$\,(\#) & method \\
   &   (K)   &       &  km s$^{-1}$ &                        &                         &      \\
\hline
1219 & 4400 & 1.90 & 1.25 & +0.19$\pm$0.08 (46)& +0.18$\pm$0.17 (15)&  $T_{\rm{eff}}$ and $\xi$ from FeII \\
36673 & 7450 & 1.90 & 4.70 & 0.00$\pm$0.16 (32)& 0.00$\pm$0.07 (12)& H$\alpha$ and $\xi$ from FeII \\
38713 & 5100 & 2.45 & 1.63 & +0.05$\pm$0.08 (62)& +0.06$\pm$0.07 (18)& H$\alpha$ and $\xi$ from FeII \\
44362 & 5600 & 1.55 & 3.09 & +0.10$\pm$0.09 (32)& +0.09$\pm$0.26 (07)& H$\alpha$ and $\xi$ from FeII \\
45348 & 7450 & 2.10 & 3.30 & $-$0.04$\pm$0.21 (35)& $-$0.13$\pm$0.05 (13)& H$\alpha$ and $\xi$ from FeII \\
49068 & 4625 & 2.20 & 1.78 & +0.19$\pm$0.08 (47) & +0.19$\pm$0.17 (17) &  $T_{\rm{eff}}$ and $\xi$ from FeI \\
49396 & 5350 & 2.15 & 5.03 & +0.14$\pm$0.11 (26) & +0.13$\pm$0.03 (06) & H$\alpha$ and $\xi$ from FeII \\
51043 & 4900 & 1.85 & 2.74 & +0.02$\pm$0.11 (34) & +0.02$\pm$0.08 (12) & H$\alpha$ and $\xi$ from FeII \\
66190 & 4785 & 1.85 & 2.67 & +0.26$\pm$0.10 (34) & +0.25$\pm$0.07 (14) &  $T_{\rm{eff}}$ and $\xi$ from FeI \\
71181 & 5100 & 2.30 & 2.59 & +0.14$\pm$0.14 (55) & +0.13$\pm$0.06 (12) & H$\alpha$ and $\xi$ from FeII \\
76860 & 4375 & 1.75 & 2.67 & +0.17$\pm$0.13 (38) & +0.17$\pm$0.18 (13) &  $T_{\rm{eff}}$ and $\xi$ from FeI \\
80404 & 7500 & 2.40 & 2.35 & $-$0.14$\pm$0.18 (26) & 0.00$\pm$0.06 (15) & H$\alpha$ and $\xi$ from FeII \\
90289 & 4100 & 1.70 & 1.49 & +0.09$\pm$0.16 (55) & +0.08$\pm$0.15 (08) &  $T_{\rm{eff}}$ and $\xi$ from FeI \\
102839 & 4670 & 1.10 & 2.80 & +0.11$\pm$0.08 (30) & +0.11$\pm$0.12 (10) &  $T_{\rm{eff}}$ and $\xi$ from FeI \\
114792 & 5600 & 2.35 & 7.44 & +0.06$\pm$0.17 (32) & +0.05$\pm$0.07 (10) & H$\alpha$ and $\xi$ from FeII \\
159633 & 5200 & 1.85 & 4.45 & +0.13$\pm$0.09 (37) & +0.14$\pm$0.06 (11) & H$\alpha$ and $\xi$ from FeII \\
192876 & 5300 & 2.20 & 3.18 & +0.22$\pm$0.08 (35) & +0.21$\pm$0.06 (09) & H$\alpha$ and $\xi$ from FeII \\
204867 & 5700 & 2.05 & 4.29 & +0.12$\pm$0.10 (38) & +0.11$\pm$0.13 (12) & H$\alpha$ and $\xi$ from FeII \\
225212 & 4100 & 0.75 & 2.95 & +0.10$\pm$0.20 (26) & +0.11$\pm$0.22 (13) &  $T_{\rm{eff}}$ and $\xi$ from FeI \\
\noalign{\smallskip}
\hline
\end{tabular}
\end{table*}

\subsection{Uncertainties of the parameters}

In order to estimate the uncertainties of the parameters, we divided the stars in two groups,  the first for stars where the T$_{\rm{eff}}$ is from H$\alpha$ fitting and $\xi$ from the FeII lines and the second for stars whose parameters were all calculated from the FeI lines. 
We then chose a representative star, with the parameters close to the mean ones of its group, and calculated the uncertainties using them. The uncertainties thus calculated were then extended to the entire group. The chosen stars are \object{HD 49396} and HD 76860. 
\begin{table}
\caption{Uncertainties in the adopted atmospheric parameters.}
\label{tab:uncer}
\centering
\begin{tabular}{c c c c}
\noalign{\smallskip}
\hline\hline
\noalign{\smallskip}
HD & T$_{\rm{eff}}$ & log g & $\xi$ \\
\hline
49396 & $\pm$200K & $\pm$0.25dex & $\pm$0.35km s$^{-1}$ \\
76860 & $\pm$200K & $\pm$0.40dex & $\pm$0.20km s$^{-1}$ \\
\noalign{\smallskip}
\hline
\end{tabular}
\end{table}
\par When determining the T$_{\rm{eff}}$ from the FeI lines, we searched for a linear fit where the angular coefficient is null. Obviously this coefficient has a statistical uncertainty. In order to find the 1$\sigma$ uncertainty on the T$_{\rm{eff}}$ determination we changed the temperature until the angular coefficient of the linear fit matched its own uncertainty. A similar procedure was followed to find the uncertainty on the $\xi$ determination. The uncertainties thus calculated are listed in Table \ref{tab:uncer}.
\par In order to find the uncertainty of the T$_{\rm{eff}}$ in the star HD 49396, we had to follow a different procedure. For this star the T$_{\rm{eff}}$ was determined from the H$\alpha$ fitting. We then changed the temperature until the calculated fit marginally agreed with the observed profile. In this sense, any fit with temperature between these limits could be considered somewhat reasonable.
\par In order to find the 1$\sigma$ uncertainty of the surface gravity, we proceeded as follows. The mean abundance as given from the FeI lines and from the FeII lines have, in general, different standard deviations. The uncertainties here are considered to be the standard deviations. We then changed the gravity until the difference between the FeI and FeII means equalled the higher standard deviation. We consider that to be the 1$\sigma$ uncertainty in log g. All the uncertainties are listed in Table \ref{tab:uncer}.

\subsection{Comparison with the literature}

Some of the stars analyzed in this work have atmospheric parameters that are published in the literature. Our determinations show an overall good agreement with them. Some values from the literature are listed in Table \ref{tab:comp} along with ours for comparison. Most of these  results are from high resolution spectroscopic analysis.
\par Out of our sample, Canopus (HD 45348) is probably the most extensively studied star. Among the published parameters, we believe the set by Jerzykiewicz \& Molenda-Zacowicz (\cite{JM00}) to be the most reliable. In that work the temperature is derived from measurements of the angular diameter and the total absolute flux. The gravity is derived by placing the star in a theoretical evolutionary diagram using the above temperature and luminosity obtained from the Hipparcos parallax and the total flux. 
\par Our temperature for Canopus is in excellent agreement with theirs, as well as with the others, as shown in Table \ref{tab:comp}. Our gravity is slightly higher but is also in agreement with theirs within the uncertainties. However it is important to stress that it is not possible to fit the observed H$\alpha$ wings with a temperature around 7500K and a smaller gravity than 2.10 dex. The other parameters also agree well with the values from the literature.
\par The picture is the same for the other stars as there is generally good agreement. However our gravity tends to be higher than previously determined, especially in the case of HD 80404. Again it is not possible to fit the observed H$\alpha$ wings for this star with a temperature around 7500K and a smaller gravity than 2.40 dex.
\par The only star for which the agreement is not good is HD 204867. Luck (\cite{L77}) employed curves of growth to do his analysis. Foy (\cite{F81}) reanalyzed the same star using equivalent widths but adopting the same temperature determined by Luck (\cite{L77}). The interesting fact, however, is that both analyses made use of the models by Gustafsson et al. (\cite{G75}). Thus the differences are mainly due to the different set of \emph{gfs} employed. This shows the importance of well-determined \emph{gfs}. It seems that our method of determining the temperature is more reliable and shows excellent agreement with the temperatures from the FeI and FeII excitation equilibria.
%
\begin{table}
\caption{Atmospheric parameters available in the literature in comparison with the present results.}
\label{tab:comp} 
\begin{tabular}{c c c c r c}
\noalign{\smallskip}
\hline\hline
\noalign{\smallskip}
HD & T$_{\rm{eff}}$ & log g & $\xi$ & [Fe/H] & Ref. \\
\hline
45348 & 7500 &    2.10     &    3.30    & $-$0.04 & this work \\
45348 & 7464 &  1.68-1.76  &    --     &   --  & (1) \\
45348 & 7575 &  1.90-2.10  &    3.00    & $-$0.25 & (2) \\
45348 & 7500 &    1.50     &    2.50    & +0.06 & (3) \\
45348 & 7500 &    1.20     &    3.00    & +0.08 & (4) \\
45348 & 7500 &    1.50     &    3.50    & $-$0.07 & (5) \\
36673 & 7450 & 1.90 & 4.70 &  0.00 & this work \\
36673 & 7350 & 1.80 & 3.00 & $-$0.05 & (2) \\
36673 & 7400 & 1.50 & 5.90 & $-$0.06 & (6) \\
36673 & 7000 & 1.30 & 2.50 & $-$0.10 & (5) \\
80404 & 7500 & 2.40 & 2.35 & $-$0.14 & this work \\
80404 & 7500 & 1.60 & 2.20 & +0.02 & (7) \\
80404 & 7500 & 0.90 & 2.50 & +0.06 & (5) \\
49068 & 4625 & 2.20 & 1.78 & +0.19 & this work \\
49068 & 4500 & 2.00 & 2.00 &  0.00 & (8) \\
204867 & 5700 & 2.05 & 4.29 & +0.12 & this work \\
204867 & 5362 & 1.15 & 3.50 & $-$0.05 & (9) \\
204867 & 5475 & 1.60 & 3.10 & $-$0.02 & (10) \\
204867 & 5475 & 1.30 & 2.30 & +0.19 & (11)  \\
225212 & 4100 & 0.75 & 2.95 & +0.10 & this work \\
225212 & 4250 & 0.80 & 4.50 & $-$0.20 & (12) \\
\noalign{\smallskip}
\hline
\end{tabular}
\\
(1) Jerzykiewicz \& Molenda-Zacowicz \cite{JM00}, (2) Luck et al. \cite{L98}, (3) Hill et al. \cite{H95}, (4) Spite et al. \cite{S89}, (5) Luck \& Lambert \cite{LL85}, (6) Venn \cite{V95a}, (7) Luck \& Lambert \cite{LL92}, (8) Gilroy \cite{G89}, (9) Luck \cite{L82}, (10) Foy \cite{F81}, (11) Luck \cite{L77}, (12) Luck \& Bond \cite{LB80}. \\
\end{table}

\subsection{Masses}

We also estimated the masses of our sample stars. To do so we placed the stars in the HR diagram with theoretical evolutionary tracks. Luminosities were calculated using  M$_V$ = V - A$_v$ + 5 +5log $\pi$ and the bolometric corrections from Alonso et al.\ (\cite{Al99}) and adopting a solar bolometric magnitude, M$_{{Bol}_{\odot}}$ = 4.75 (Cram \cite{C99}), $log(L_{\star} / L_{\odot}) = -0.4(M_{{Bol}_{\star}} - M_{{Bol}_{\odot}})$.
\begin{figure}
\resizebox{\hsize}{!}{\includegraphics{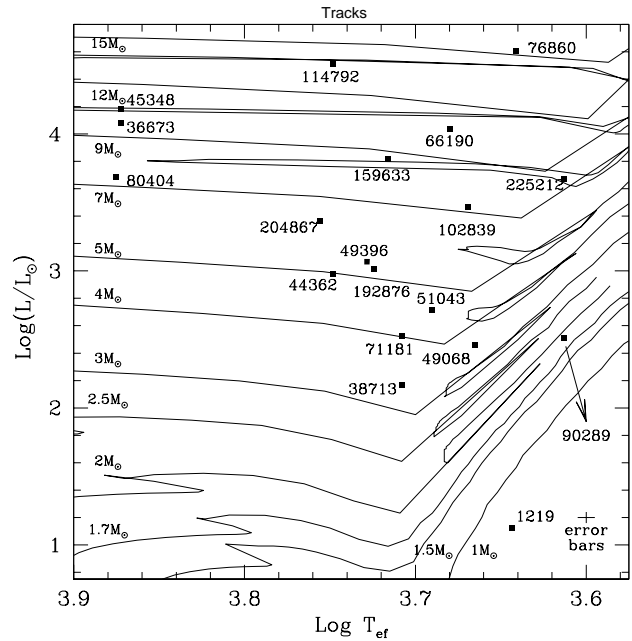}}
\caption{The HR diagram with our stars and the theoretical evolutionary tracks from Schaller et al.\ (\cite{Sc92}).}
\label{fig:hr}
\end{figure}
\par Once the stars are placed in the HR diagram their masses can be estimated by interpolating among the tracks. Some of the stars fall in regions where blue loops may occur, so they have two mass estimates, one for the first crossing and the second one for the blue loop. These masses are listed in Table \ref{tab:mass}.
\begin{table}
\caption{The bolometric magnitudes, bolometric corrections, luminosities, and estimated masses.}
\centering 
\label{tab:mass}
\begin{tabular}{c c c c c}
\noalign{\smallskip}
\hline\hline
\noalign{\smallskip}
 HD & M$_{Bol}$ & BC & log (L$_{\star}$/L$_{\odot}$) & Mass \\
    &           &    &                               & in M$_{\odot}$ \\
\hline
1219 & 1.94 & -0.57 & 1.13 & -- \\
36673 & -5.46 & 0.00 & 4.08 & 10.1 - 8.6 \\
38713 & -0.67 & -0.23 & 2.17 & 3.4 \\
44362 & -2.70 & -0.12 & 2.98 & 5.0 \\
45348 & -5.71 & 0.00 & 4.18 & 10.6 - 9.0 \\
49068 & -1.41 & -0.42 & 2.46 & 3.5 \\
49396 & -2.92 & -0.16 & 3.07 & 5.4 \\
51043 & -2.04 & -0.30 & 2.72 & 4.6 \\
66190 & -5.34 & -0.34 & 4.04 & 10.6 - 8.3 \\ 
71181 & -1.57 & -0.23 & 2.53 & 4.0 \\
76860 & -6.77 & -0.58 & 4.62 & 15.5 \\
80404 & -4.47 & -0.01 & 3.69 & 7.4 \\
90289 & -1.52 & -0.83 & 2.51 & 1.9 \\ 
102839 & -3.91 & -0.39 & 3.46 & 7.3 - 6.0 \\
114792 & -6.53 & -0.12 & 4.51 & 13.8 - 11.8 \\
159633 & -4.80 & -0.20 & 3.82 & 8.8 - 7.0 \\
192876 & -2.79 & -0.17 & 3.02 & 5.3 \\ 
204867 & -3.66 & -0.10 & 3.37 & 6.5 - 6.0 \\
225212 & -4.42 & -0.83 & 3.67 & 7.9 - 7.0  \\
\noalign{\smallskip}
\hline
\end{tabular}
\end{table}
 HD 1219 falls bellow the tracks, so its mass could not be estimated. Probably its parallax is wrong, leading to wrong distance and wrong reddening and luminosity. \object{HD 90289} seems to be in the AGB region of the HR diagram.
\par Based on the uncertainties of the parallaxes, visual magnitudes, extinctions, and bolometric corrections, we estimate the mean uncertainty of log (L$_{\star}$/L$_{\odot}$) to be $\approx$ 0.08 dex. The mean uncertainty of log (T$_{\rm{eff}}$) is $\approx$ 0.01 dex. There are, however, other sources of uncertainties, which we cannot estimate, affecting the luminosities. First, the evolutionary tracks we used are for solar metallicity stars, however, most of our sample stars are slightly more metallic than that. Second, the adopted tracks do not take rotation into account. Rotation is supposed to change not only the photospheric abundances but also the evolutionary path along the HR diagram.


\section{Abundances}

In this section we discuss the determination of CNO abundances. All the abundances were derived using spectral synthesis. The codes for calculating synthetic spectra are described by Barbuy et al.\ (\cite{B03}). The adopted C, N, and O atomic lines are listed in Table \ref{tab:cnolines} with corresponding excitation potential and oscillator strength.

\begin{table}
\caption{Data of the adopted C, N, and O atomic lines.}
\centering 
\label{tab:cnolines}
\begin{tabular}{c c c c}
\noalign{\smallskip}
\hline\hline
\noalign{\smallskip}
 Species & $\lambda$ (\AA) & $\chi$ (eV)  & log $gf$ \\
\hline
CI & 5380.322 & 7.68 & -1.640 \\
NI & 7442.310 & 10.33 & -0.385 \\ 
NI & 7468.312 & 10.33 & -0.190 \\
NI & 8200.357 & 10.33 & -1.001 \\
NI & 8210.715 & 10.33 & -0.708 \\
NI & 8216.336 & 10.33 & +0.132 \\ 
NI & 8242.389 & 10.33 & -0.256 \\
OI & 6156.737 & 10.74 & -1.487 \\
OI & 6156.755 & 10.74 & -0.898 \\
OI & 6156.778 & 10.74 & -0.694 \\
OI & 6158.149 & 10.74 & -1.841 \\ 
OI & 6158.172 & 10.74 & -0.995 \\
OI & 6158.187 & 10.74 & -0.409 \\
$[$OI$]$& 6300.311 & 0.00 & -9.716 \\
\noalign{\smallskip}
\hline
\end{tabular}
\end{table}

\subsection{Carbon}

Carbon abundances were calculated from the CI line $\lambda$5380.322\AA\ for stars that are hotter than 5200K and from the C$_{2}$ lines at $\lambda$5135.62\AA\ for cooler stars. The oscillator strength of the CI line, log $gf$ = $-$1.64, was derived by fitting the solar spectrum with the solar abundance recommended by Grevesse \& Sauval (\cite{G98}), A$_C$ = 8.52. We used the solar spectrum available on the internet\footnote{The spectrum is freely available for download at the ESO website: www.eso.org/observing/dfo/quality/UVES/pipeline/solar\_spectrum.html} observed with UVES at the VLT. 
\par
\begin{figure}
\resizebox{\hsize}{!}{\includegraphics{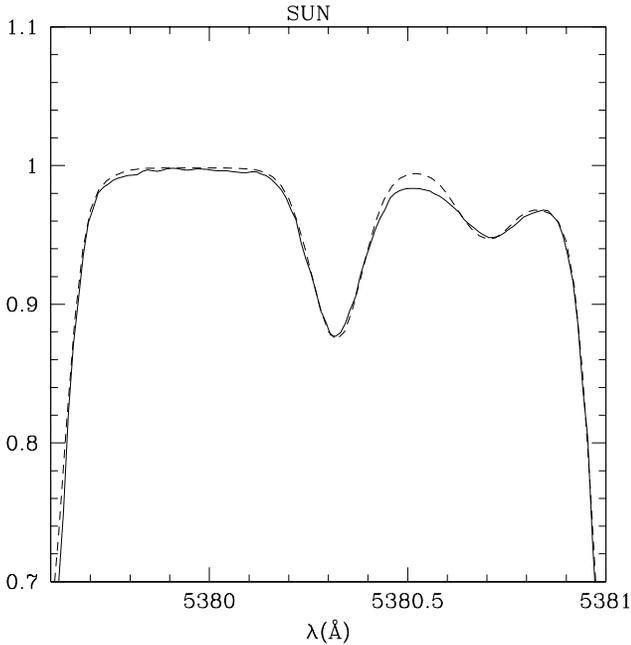}}
\caption{Fit to the CI $\lambda$5380.320\AA\ line in the Sun. The synthetic spectrum (dashed) is compared to the observed one (solid).}
\label{fig:car1}
\end{figure}

\begin{figure}
\resizebox{\hsize}{!}{\includegraphics{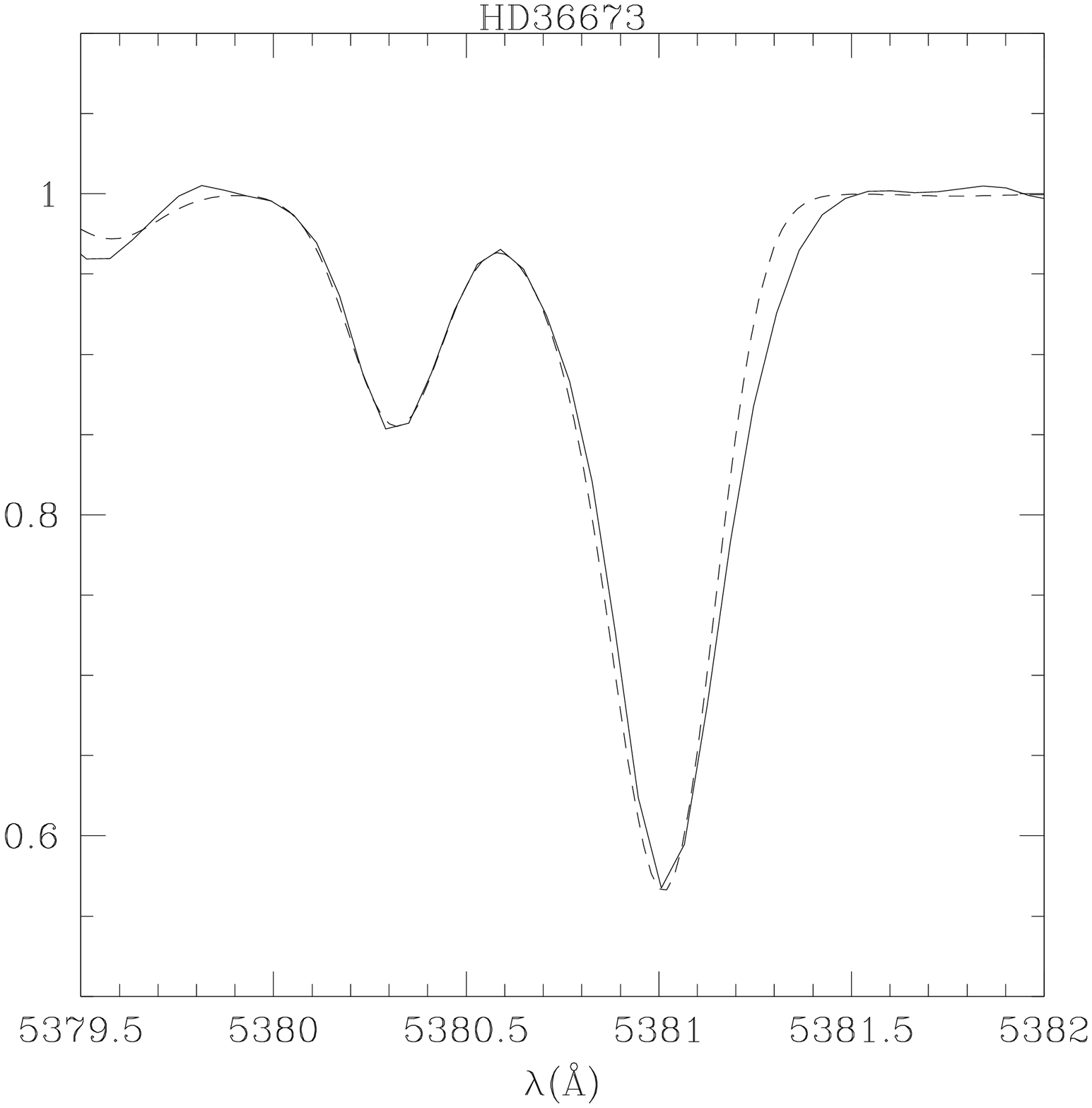}}
\caption{Fit to the CI $\lambda$5380.320\AA\ line in HD 36673. The synthetic spectrum (dashed) is compared to the observed one (solid).}
\label{fig:car2}
\end{figure}

 A blend of lines on the red side of the CI line was treated as a set of FeI lines following Spite et al.\ (\cite{S89}). The solar model was constructed using the grids of Kur\'ucz (\cite{K94}) and the parameters T$_{\rm{eff}}$ = 5780K, log g = 4.44dex, and $\xi$ = 1.00km s$^{-1}$. Figures \ref{fig:car1} and \ref{fig:car2} show the observed spectrum and the synthetic fits for the Sun and the star HD 36673.
\par
\begin{figure}
\resizebox{\hsize}{!}{\includegraphics{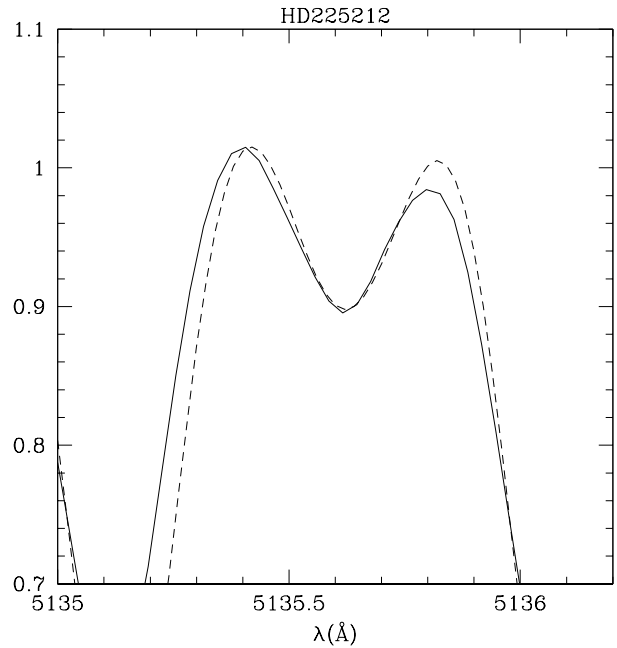}}
\caption{Fit to the C$_2$ $\lambda$5135.62\AA\ line in HD 225212. The synthetic spectrum (dashed) is compared to the observed one (solid).}
\label{fig:c2}
\end{figure}

  The C$_2$(0,0) $\lambda$5135.62\AA\ is a band of the Swan system. The data of the C$_2$ molecule are those by Barbuy (\cite{B85}), dissociation potential D$_0$(C$_2$) = 6.21 eV and electronic-vibrational oscillator strength $f_{00}$ = 0.0184. An example of fit is shown in Fig. \ref{fig:c2} for \object{HD 225212}.
\par Venn (\cite{V95a}) analyzed the influence of NLTE effects in the carbon abundances derived from CI lines in a sample of A0-F0 supergiants. It was shown that NLTE is important, and the abundances derived by assuming ETL must be corrected. The amplitude of the correction increases from F0 type stars to A0 type.
\par Even though we do not have the means to estimate the exact correction that must be applied, we adopted a mean correction. Among the stars analyzed by Venn (\cite{V95a}), HD 36673 is also in our sample. There are also two other stars with similar spectral type, \object{HD 25291} (F0II) and \object{HD 6130} (F0II). Venn's analysis has shown that the mean carbon abundances of these stars should be corrected by $-$0.25dex (HD 36673), $-$0.24dex (HD 25291), and $-$0.16dex (HD 6130). We adopted the mean value, $-$0.22dex, as the correction that must be applied to the carbon abundances in the stars HD 36673 (F0Ib) and HD 45348 (F0II).
\par The abundance of the star HD 80404 (A8Ib) also needs to be corrected. For a star with similar spectral type, \object{HD 58585} (A8II), Venn (\cite{V95a}) adopts a correction of $-$0.33dex. We adopted the same value for HD 80404. The carbon abundances (as well as the nitrogen and oxygen abundances) are listed in Table \ref{tab:abun}. The abundances listed have already been corrected for NLTE effects whenever necessary.

\subsection{Nitrogen}

Nitrogen abundances were derived from atomic lines for hotter stars than 5200K and from CN molecular lines for the cooler stars. We used two atomic lines from the multiplet 3 around $\lambda$7444\AA\ and four lines of the multiplet 2 around $\lambda$8220\AA. The line list and the $gfs$ are listed in Table \ref{tab:cnolines}. The adopted $gfs$ are the ones recommended by NIST (Martin et al. \cite{NIST}).
\par The CN lines we used are the CN(5,1) $\lambda$6332.18\AA\, and CN(6,2) $\lambda$6478.48\AA\ bandheads of the A$^2\Pi$-X$^2\Sigma$ red system. The data of the CN lines are the same as adopted by Milone et al.\ (\cite{M92}), dissociation potential D$_0$(CN) = 7.65 eV and electronic oscillator strength $f_{el}$ = 6.76 10$^{-3}$. We adopted the solar abundance of nitrogen recommended by Grevesse \& Sauval (\cite{G98}), A$_N$ = 7.92.
\par
\begin{figure}
\resizebox{\hsize}{!}{\includegraphics{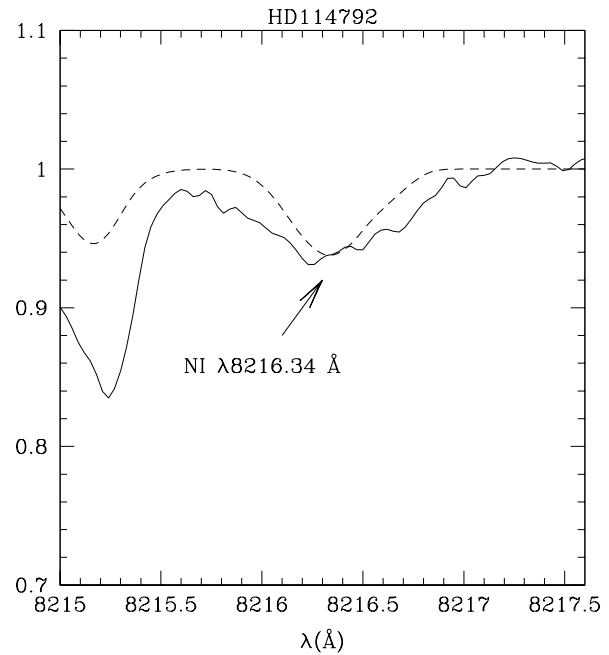}}
\caption{Fit to the NI $\lambda$8216.34\AA\ in HD 114792 line. The synthetic spectrum (dashed) is compared to the observed one (solid). This star has a temperature of about 5600K. The nitrogen line is weak and affected by many weak unidentified blends.}
\label{fig:nit1}
\end{figure}

 The region around $\lambda$8220\AA\ is highly contaminated by telluric lines. In order to  properly identify the telluric lines, we carefully compared the spectra of stars with distinct radial velocities. Any nitrogen line blended with a telluric one was excluded from the analysis. Since weak telluric lines may not be properly identified, some of the nitrogen lines may be slightly contaminated. Moreover, most of our stars are not hot enough to allow the high excitation atomic nitrogen lines to be well defined. They are also affected by some weak unidentified blends. Thus the synthetic fit for these lines should be considered with care, possibly as an upper limit for the abundance. Figure \ref{fig:nit1} exemplifies this situation.
\par
\begin{figure}
\resizebox{\hsize}{!}{\includegraphics{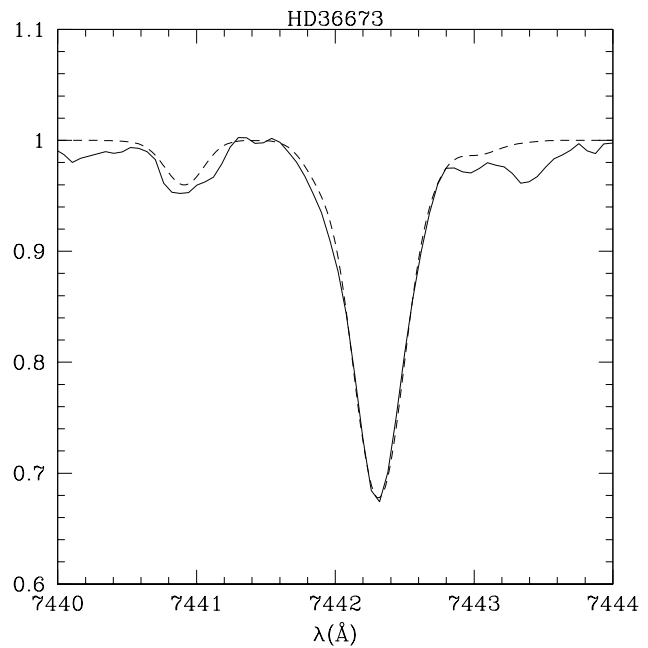}}
\caption{Fit to the NI $\lambda$7442.31\AA\ line in HD 36673. The synthetic spectrum (dashed) is compared to the observed one (solid).}
\label{fig:nit36673}
\end{figure}

\begin{figure}
\resizebox{\hsize}{!}{\includegraphics{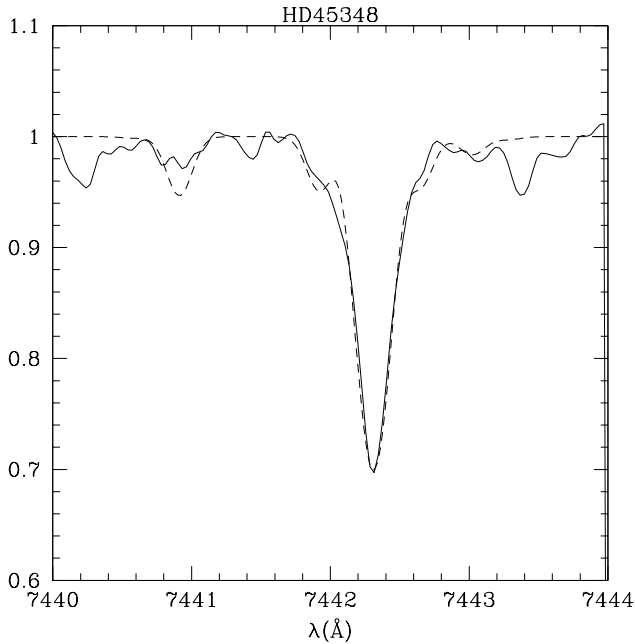}}
\caption{Fit to the NI $\lambda$7442.31\AA\ line in HD 45348. The synthetic spectrum (dashed) is compared to the observed one (solid).}
\label{fig:nit45348}
\end{figure}

 The two lines in the region around $\lambda$7440\AA\ have no problem with telluric lines but are also affected by unidentified blends. We tried to simulate the blends with FeI lines for the three hottest stars. In these stars there are blends in both wings of both the adopted lines. These artificial lines did a good job in adjusting the line wings of the stars HD 36673 and HD 80404, but not as good for HD 45348, as shown in Figs. \ref{fig:nit36673} and \ref{fig:nit45348}.
\par
\begin{figure}
\resizebox{\hsize}{!}{\includegraphics{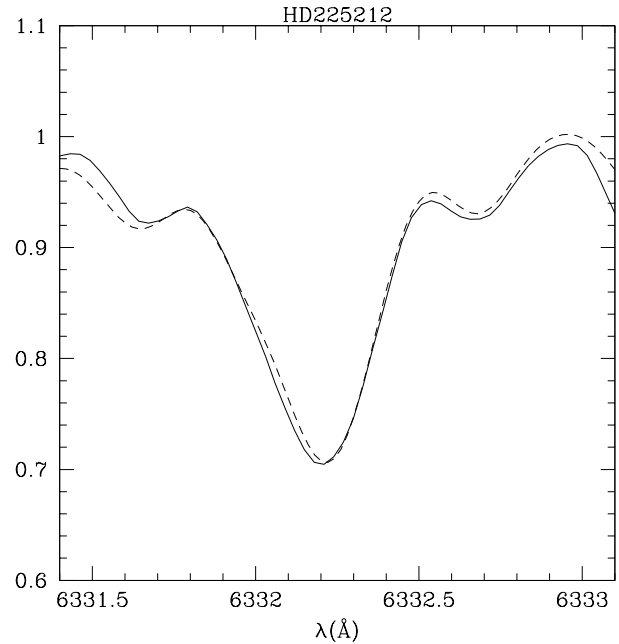}}
\caption{Fit to the CN band $\lambda$6332.18\AA\ in HD 225212. The synthetic spectrum (dashed) is compared to the observed one (solid).}
\label{fig:CN51}
\end{figure}

\begin{figure}
\resizebox{\hsize}{!}{\includegraphics{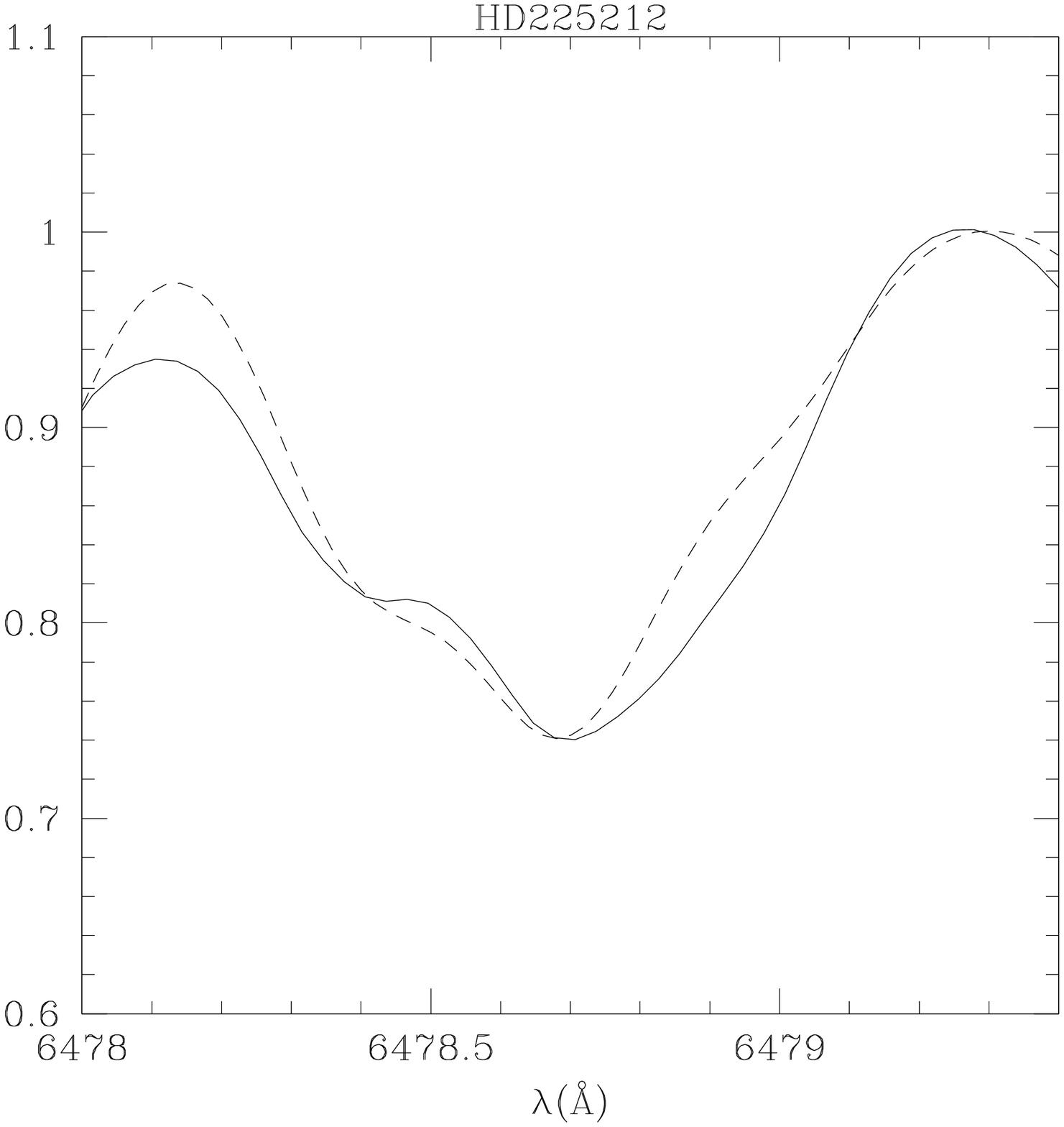}}
\caption{Fit to the CN band $\lambda$6478.48\AA\ in HD 225212. The synthetic spectrum (dashed) is compared to the observed one (solid).}
\label{fig:CN62}
\end{figure}

 The molecular CN lines are also affected by blends. The band at $\lambda$6332.18\AA\ is affected by two lines at $\lambda$6331.95\AA, one due to SiI and the other due to FeII. Both are taken into account in the synthesis. In general the fit to this CN line is better than the fit to the $\lambda$6478.48\AA\ line. Figures \ref{fig:CN51} and \ref{fig:CN62} are examples of the fits for the star HD 225212.
\par The nitrogen abundances of the stars HD 36673, HD 45348, and HD 80404 are affected by NLTE. Venn (\cite{V95a}) investigated the effects of NLTE in the nitrogen abundances. In order to correct our results from NLTE effects, we proceeded as we did for carbon, by adopting mean corrections based on the results by Venn (\cite{V95a}) for stars with similar spectral type to ours. Thus, the mean nitrogen abundances were corrected by $-$0.31dex for HD 36673 and HD 45348 and by $-$0.58dex for HD 80404. Table \ref{tab:nitline} lists the abundances derived from each line for each star. In this table the abundances have not been corrected for NLTE, the ones  corrected for NLTE are listed in Table \ref{tab:abun}.
\begin{table*}
\caption{The abundances of nitrogen line by line in each star. The abundances in this table are not corrected for NLTE.}
\centering 
\label{tab:nitline}
\begin{tabular}{c c c c c c c c c}
\noalign{\smallskip}
\hline\hline
\noalign{\smallskip}
 HD & 7442 & 7468 & 8200 & 8210 & 8216 & 8242 & CN(5,1) & CN(6,2) \\
\hline
 1219 & -- & -- & -- & -- & -- & -- & 8.07 & 7.92 \\
 36673 & 8.88 & 9.01 & 8.67 & -- & -- & -- & -- & -- \\
 38713 & -- & -- & -- & -- & -- & -- & 8.27 & -- \\
 44362 & 8.24 & -- & -- & -- & -- & -- & -- & -- \\
 45348 & 8.60 & 8.72 & -- & 8.60 & -- & -- & -- & -- \\
 49068 & -- & -- & -- & -- & -- & -- & 8.63 & 8.71 \\
 49396 & 8.74 & -- & -- & -- & -- & -- & -- & -- \\
 51043 & -- & -- & -- & -- & -- & -- & 8.54 & -- \\
 66190 & -- & -- & -- & -- & -- & -- & 9.01 & 9.15 \\
 71181 & -- & -- & -- & -- & -- & -- & 8.84 & 8.75 \\
 76860 & -- & -- & -- & -- & -- & -- & 8.92 & 8.80 \\
 80404 & 8.82 & 9.06 & -- & 8.68 & -- & 8.62 & -- & -- \\
 90289 & -- & -- & -- & -- & -- & -- & -- & -- \\
 102839 & -- & -- & -- & -- & -- & -- & 8.46 & 8.37 \\
 114792 & 8.30 & 8.27 & -- & -- & 8.19 & 8.46 & -- & -- \\
 159633 & -- & -- & -- & -- & -- & -- & -- & -- \\
 192876 & 8.66 & -- & -- & -- & -- & -- & -- & -- \\
 204867 & 8.58 & 8.24 & -- & -- & -- & -- & -- & -- \\
 225212 & -- & -- & -- & -- & -- & -- & 8.47 & 8.39 \\
\noalign{\smallskip}
\hline
\end{tabular}
\end{table*}

\subsection{Oxygen}

Oxygen abundances were calculated from two lines of the OI permitted triplet, at  $\lambda$6156.7\AA\ and $\lambda$6158.1\AA\, for the three hottest stars, and from the [OI] forbidden line, at $\lambda$6300.311\AA\, for the other stars. We adopted the recommended data for the fine structure of the permitted lines from NIST (Martin et al. \cite{NIST}). The atomic data used are reported in Table \ref{tab:cnolines}. The solar abundance we adopted is the one suitable for the 1D models recommended by Allende Prieto et al.\ (\cite{AP01}), A$_{O}$ = 8.77.
\par
\begin{figure}
\resizebox{\hsize}{!}{\includegraphics{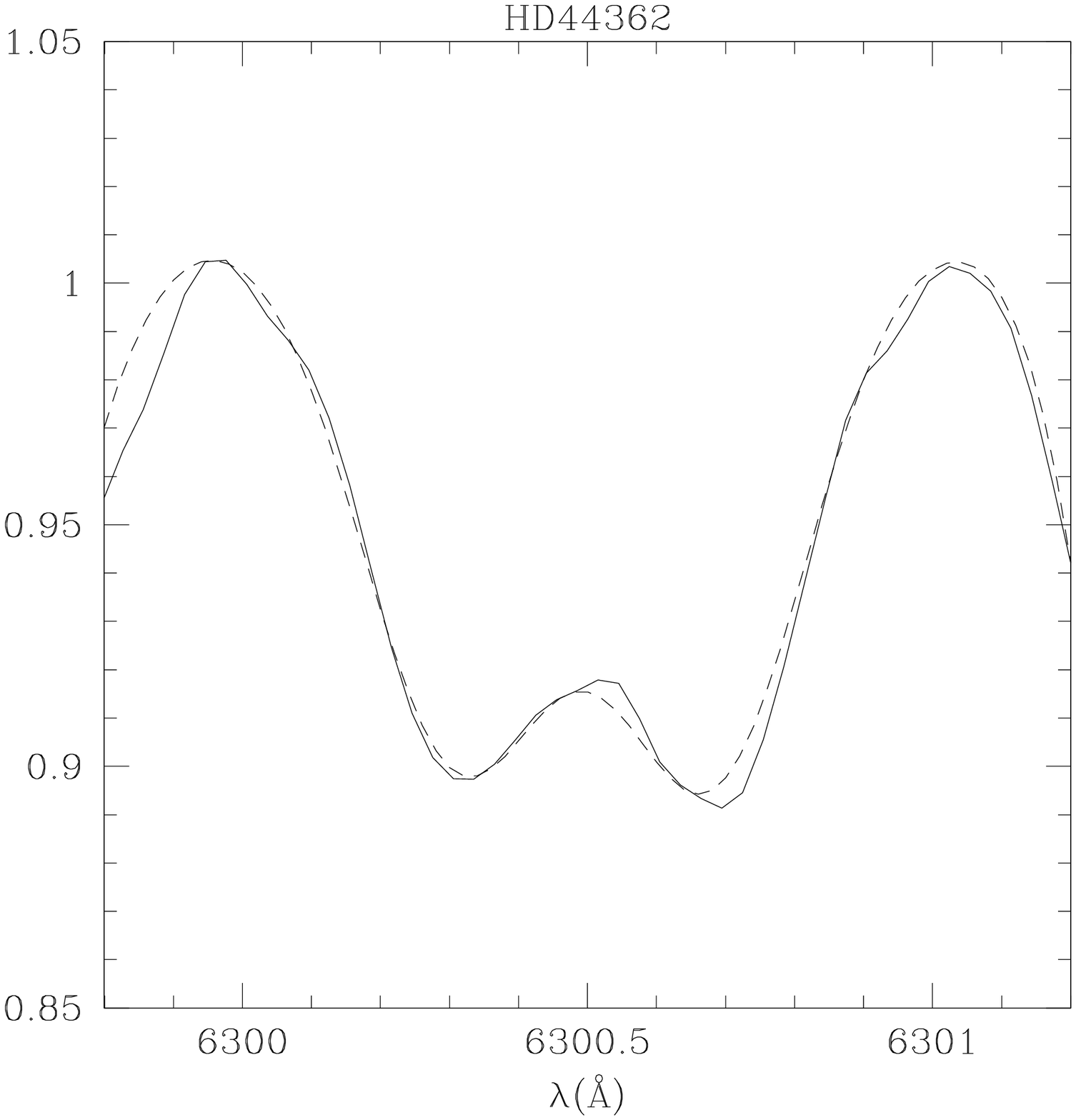}}
\caption{Fit to the [OI] $\lambda$6300.311\AA\ line in HD 44362. The synthetic spectrum (dashed) is compared to the observed one (solid).}
\label{fig:for1}
\end{figure}

\begin{figure}
\resizebox{\hsize}{!}{\includegraphics{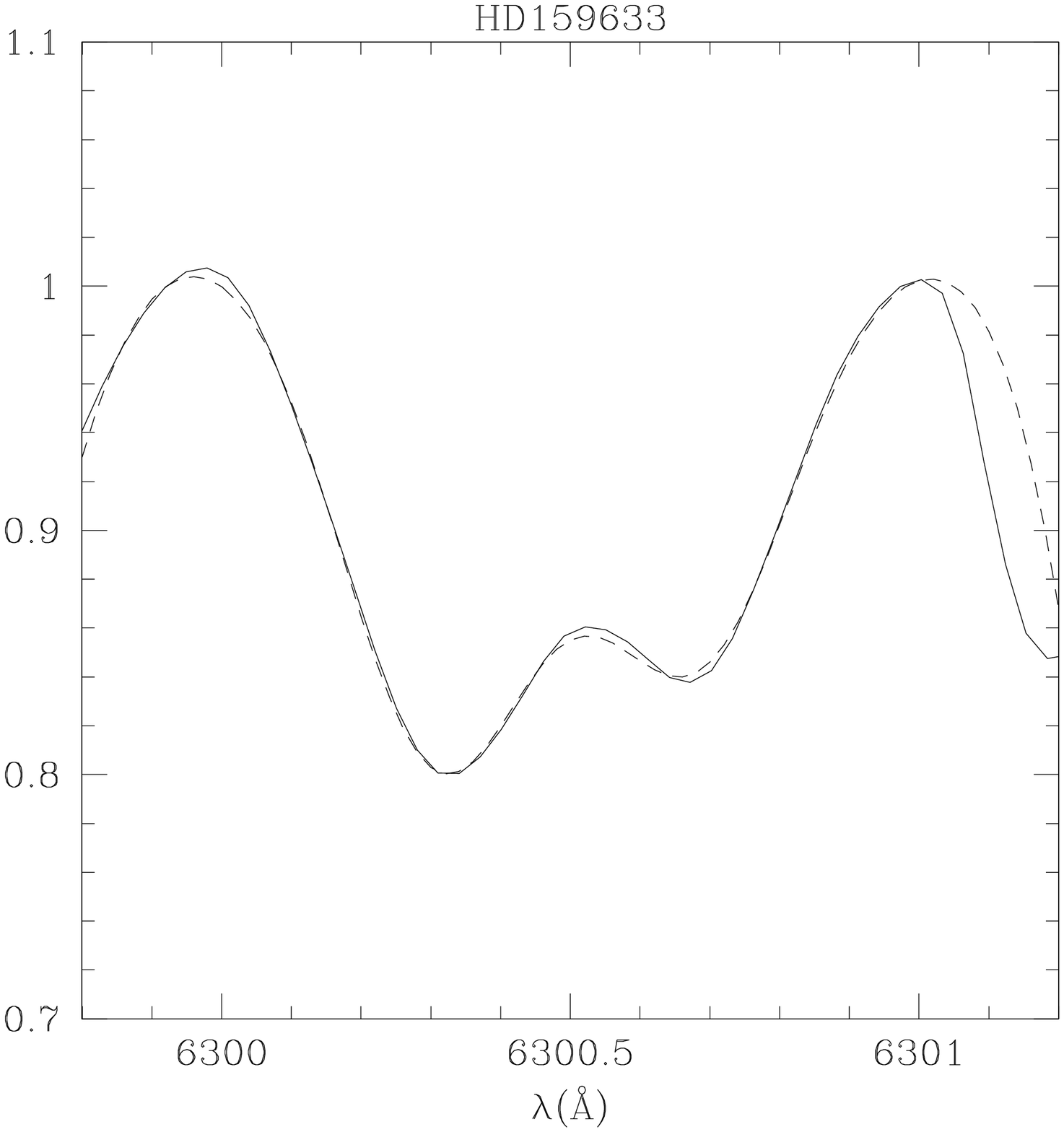}}
\caption{Fit to the [OI] $\lambda$6300.311\AA\ line in HD 159633. The synthetic spectrum (dashed) is compared to the observed one (solid).}
\label{fig:for2}
\end{figure}

\begin{figure}
\resizebox{\hsize}{!}{\includegraphics{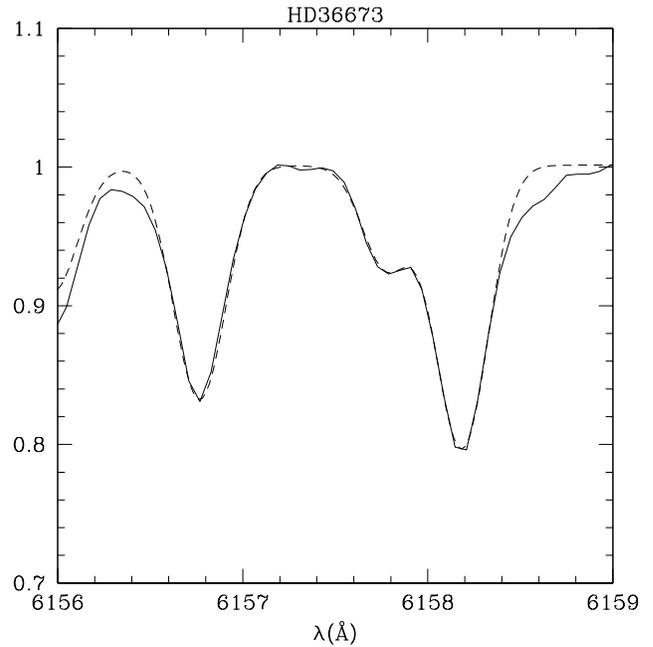}}
\caption{Fit to the OI $\lambda$6156.7\AA\ and $\lambda$6158.1\AA\ lines in HD 36673. The synthetic spectrum (dashed) is compared to the observed one (solid).}
\label{fig:per1}
\end{figure}

 The forbidden line is blended with a weak NiI line at $\lambda$6300.34\AA, which is included in the synthesis with parameters recommended by Allende Prieto et al.\ (\cite{AP01}). It also has a nearby ScII line at $\lambda$6300.70\AA\, for which we adopted the hyperfine structure by Spite et al. (\cite{S89}). In Figs. \ref{fig:for1} and \ref{fig:for2} we show examples of the fits for the forbidden line in \object{HD 44362} and \object{HD 159633}, while  Fig. \ref{fig:per1} shows an example of the fit for the permitted lines in HD 80404.
\begin{table}
\caption{Abundances of C, N, and O.}
\centering 
\label{tab:abun}
\begin{tabular}{c c c c}
\noalign{\smallskip}
\hline\hline
\noalign{\smallskip}
 HD & [C/Fe] & [N/Fe]  & [O/Fe] \\
\hline
1219   & 0.00 & $-$0.12 &    -- \\
36673  & $-$0.86 & +0.62 & $-$0.25 \\
38713  & $-$0.14 & +0.30 & $-$0.14 \\
44362  &   --    & +0.22 & $-$0.31 \\
45348  & $-$0.55 & +0.45 & $-$0.11 \\
49068  & $-$0.23 & +0.56 & +0.06 \\
49396  & $-$0.61 & +0.68 & +0.10 \\
51043  & $-$0.19 & +0.60 & $-$0.07 \\
66190  & $-$0.39 & +0.90 & 0.00 \\
71181  & $-$0.23 & +0.74 & +0.09 \\
76860  & $-$0.24 & +0.77 & +0.13 \\
80404  & $-$0.55 & +0.44 &   +0.08 \\
90289  & $-$0.13 &   --  &   +0.13 \\
102839 & $-$0.26 & +0.39 & $-$0.13 \\
114792 & $-$0.49 & +0.33 &   +0.14 \\
159633 &    --   &   --  & +0.06 \\
192876 & $-$0.61 & +0.52 & $-$0.01 \\
204867 & $-$0.63 & +0.37 & $-$0.06 \\
225212 & $-$0.14 & +0.41 & $-$0.02 \\
\noalign{\smallskip}
\hline
\end{tabular}
\end{table}
\par According to Takeda \& Takada-Hidai (\cite{TT98}), the oxygen abundances, as derived from the permitted triplet in $\lambda$6156\AA\, for stars hotter than 5750K, are affected by NLTE. This temperature limit applies to HD 36673, HD 45348, and HD 80404. For stars with similar temperatures to ours, 7500K, Takeda \& Takada-Hidai (\cite{TT98}) estimate a correction of about $-$0.15dex. We thus corrected our abundances by that amount. The oxygen abundances, corrected for NLTE whenever necessary, are listed in Table \ref{tab:abun}.
\begin{table*}
\caption{The uncertainties of the abundances.}
\centering 
\label{tab:sigma}
\begin{tabular}{c c c c c c c c}
\noalign{\smallskip}
\hline\hline
\noalign{\smallskip}
 HD  & Elem. & Line & $\sigma_{Teff}$ & $\sigma_{log g}$ & $\sigma_{\xi}$ & $\sigma_{[Fe/H]}$ & $\sigma_{total}$ \\
\hline
49396 & C & $\lambda$5380.32\AA & -0.09 & +0.15 & +0.05 & -0.06 & $\pm$0.19 \\
76860 & C & C$_2$ $\lambda$5135.62\AA\ & -0.05 & +0.06 & -0.01 & -0.07 & $\pm$0.11 \\
36673 & N & $\lambda$7442.31\AA & -0.02 & +0.03 &  0.00 & -0.16 & $\pm$0.16 \\
36673 & N & $\lambda$7468.31\AA & 0.00  & +0.05 &  0.00 & -0.16 & $\pm$0.17 \\
36673 & N & $\lambda$8200.36\AA & -0.01 & +0.04 & 0.00  & -0.16 & $\pm$0.17 \\
76860 & N & CN $\lambda$6332.18\AA & -0.04 & +0.15 & +0.02 & -0.02 & $\pm$0.16 \\
76860 & N & CN $\lambda$6478.48\AA & -0.08 & +0.02 & -0.06 & -0.13 & $\pm$0.17 \\
36673 & O & ~$\lambda$6157\AA & -0.03 & +0.03 & 0.00 & -0.17 & $\pm$0.18 \\
49396 & O & $\lambda$6300.31\AA & +0.04 & +0.11 & 0.00 & -0.07 & $\pm$0.14 \\
76860 & O & $\lambda$6300.31\AA & -0.01 & +0.10 & -0.01 & -0.08 & $\pm$0.13 \\
\noalign{\smallskip}
\hline
\end{tabular}
\end{table*}

\subsection{Uncertainties of the abundances}

The main source of uncertainties in the abundances are those in the determining the atmospheric parameters. In order to estimate the uncertainties in the abundances, we changed each atmospheric parameter by its uncertainty, keeping the other ones with the original adopted values, and recalculated the abundances. In this way one measures the effect of the parameter uncertainty in the abundance. Assuming that the effects of the uncertainties of the parameters  are independent, we can calculate the total uncertainty, $\sigma_{total}$, 
\begin{equation}
\sigma_{total} = \sqrt{(\sigma_{Teff})^2 + (\sigma_{log g})^2 + (\sigma_{\xi})^2 + (\sigma_{[Fe/H]})^2},
\label{eq:sigma}
\end{equation}
the results are listed in Table \ref{tab:sigma}.
\par Since we used the stars HD 49396 and HD 76860 to estimate the uncertainties in the parameters, the obvious choice was to choose them to estimate the uncertainties in the abundances. However the nitrogen abundance in the star HD 49396 was estimated with only one line, and that line has the same problems as were discussed above and exemplified in Fig. \ref{fig:nit1} for HD 44362. We then chose another star in order to estimate the uncertainties in the nitrogen abundances, HD 36673. We also estimated the uncertainty in the oxygen abundance as derived from the permitted lines using HD 36673, since for both HD 49396 and HD 76860, the oxygen abundance was derived from the forbidden line. As can be noted in Table \ref{tab:sigma}, $\sigma_{total}$ is always less than $\pm$0.20dex.

\section{Discussion}

In order to discuss the evolutionary status of the stars, in particular whether the first dredge-up has already happened, it is more appropriate to use the ratio [N/C] rather than solely the N or C abundances. This ratio is listed in Table \ref{tab:nc} together with the sum of the C, N, and O abundances. We exclude HD 1219 in the forthcoming discussion since we were not able to estimate its mass.
\par The mean [N/C] ratio for the sample stars is [N/C] = +0.95$\pm$0.28. The high value of the standard deviation is due to the high spread among the [N/C] values. The post first dredge-up prediction of Schaller et al.\ (\cite{Sc92}) is approximately [N/C] = +0.60 for stars between 2 and 15 M$_{\odot}$, with smaller variations than 0.05 dex. 
\par Meynet \& Maeder (\cite{MeM00}) calculate evolutionary models with an initial rotation velocity of 300 km s$^{-1}$ for 9-120M$_{\odot}$ stars. They also calculate non-rotating models in the same range of mass using the same input physics (opacities, nuclear rates, etc.). For the non-rotating models only the one with 9M$_{\odot}$ develops a blue loop. In the rotating models, the 12M$_{\odot}$ one also has a blue loop.
\par For the 9M$_{\odot}$, Meynet \& Maeder (\cite{MeM00}) predict [N/C] = +0.72 without rotation and [N/C] = +1.15 with rotation in the blue supergiant phase, after the dredge-up and during the blue loop. In the case of the 12M$_{\odot}$ model without rotation, no change is predicted in the abundances. This model has no blue loop, thus a blue supergiant with this mass is only predicted to be crossing the HR diagram before the dredge-up. The 12M$_{\odot}$ model with rotation is predicted to have [N/C] = +1.24. Less massive models are not calculated. However, it is probably reasonable to assume that the non-rotating less massive models, as well as the 12M$_{\odot}$ model, would show similar abundances to the 9M$_{\odot}$ model, [N/C] = +0.72, after the first dredge-up.
\par Adopting a tolerance of $\pm$0.20dex, [N/C] ratios between +0.52 and +0.92 would be in agreement with the Meynet \& Maeder (\cite{MeM00}) results for non-rotating stars. This means that no extra mixing process is needed to explain the abundances observed in the stars \object{HD 49068}, \object{HD 51043}, HD 102839, HD 114792, and HD 225212. Their positions in the HR diagram (Fig. \ref{fig:hr}) seem to indicate that they are post first dredge-up stars. HD 114792 is definitely along a blue loop as is probably HD 102839.
\begin{table}
\caption{[N/C] ratio and the sums C+O, C+N, and C+N+O.}
\centering 
\label{tab:nc}
\begin{tabular}{c c c c c}
\noalign{\smallskip}
\hline\hline
\noalign{\smallskip}
 HD & [N/C] & C+N & C+O & C+N+O \\
\hline
 1219  & -0.12 & 8.79 &  --  &  -- \\ 
 36673 & +1.48 & 8.59 & 8.55 & 8.85 \\
 38713 & +0.44 & 8.66 & 8.87 & 8.97 \\
 44362 &   --  &  --  &  --  & --   \\
 45348 & +1.00 & 8.48 & 8.68 & 8.84 \\
 49068 & +0.79 & 8.88 & 8.87 & 9.08 \\
 49396 & +1.29 & 8.82 & 9.06 & 9.23 \\
 51043 & +0.79 & 8.76 & 8.87 & 9.04  \\
 66190 & +1.29 & 9.16 & 9.12 & 9.40 \\
 71181 & +0.97 & 8.95 & 9.10 & 9.28 \\
 76860 & +1.01 & 9.00 & 9.16 & 9.34 \\
 80404 & +0.99 & 8.37 & 8.76 & 8.87 \\
 90289 &   --  &  --  & 9.11 &  --  \\
102839 & +0.65 & 8.70 & 8.90 & 9.03 \\
114792 & +0.82 & 8.51 & 9.02 & 9.10 \\
159633 &   --  &  --  &  --  &  --  \\
192876 & +1.13 & 8.77 & 9.04 & 9.19 \\
204867 & +1.00 & 8.56 & 8.89 & 9.01 \\
225212 & +0.55 & 8.76 & 9.00 & 9.11 \\
\noalign{\smallskip}
\hline
\end{tabular}
\end{table}
\par The same argument could be used to argue that HD 38713 would not have passed through the first dredge-up, although its abundances are definitely altered. Thus, only an earlier mixing episode could be responsible for its abundances. However, if the same $\pm$0.20dex tolerance is adopted for the Schaller et al.\ (\cite{Sc92}) results, the [N/C] ratio of HD 38713 would be in marginal agreement with the predictions. Thus, we cannot be completely sure of its status.
\par There are four stars, HD 36673, HD 49396, HD 66190, and \object{HD 192876}, with [N/C] higher than +1.10dex. These abundances cannot be explained by the non-rotating models. HD 36673 and HD 66190 are in blue loops. Their masses are around 8.5M$_{\odot}$ but they seem to be more mixed than  predicted for the 9M$_{\odot}$ rotating model, especially HD 36673.
\par The abundances in HD 49396 and HD 192876 are a little more difficult to understand. In the HR diagram (Fig. \ref{fig:hr}), they are placed where changed abundances would not be expected. The predicted blue loops for these lower masses do not extend this far. 
\par Uncertainties in the abundances alone cannot explain this picture. In order to be unmixed stars, the [N/C] would need to be wrong by more than 1.0dex. It is very unlikely for this to be the case. Even though we consider these abundances to be highly uncertain, they at least indicate the stars are fully mixed and thus they are probably post first dredge-up stars.
\par If we consider the effective temperatures and luminosities to be right, only a more extended blue loop for these low masses would reconcile the tracks and the abundances. A T$_{\rm{eff}}$ that is reduced by its uncertainty, 200K, would not bring the results to any agreement. However, the problem could lie in the luminosities.
\par As discussed before, at least for one star, HD 114792, we had a clear indication of problems with the A$_V$ calculated from the work of Hakkila et al.\ (\cite{HK97}). For both HD 49396 and HD 192876, we also adopted A$_V$ from that work. However, whereas for HD 114792 it seemed to be too high, for HD 49396 and HD 192876 it could be too low. 
\par For both stars the photometric temperatures are smaller than the temperatures deduced from H$\alpha$. This would indicate that the reddening correction is underestimated. An increase in A$_V$ would cause the stars to have a higher photometric temperature and higher luminosity. A higher luminosity is just what is needed to reconcile the position of the stars in the HR diagram to the blue loops of slightly more massive stars. However if one calculates the A$_V$ needed to bring the temperatures into agreement, we see that it is not sufficient to cause an appreciable change in the luminosity. We are thus led to believe these stars are post first dredge-up stars in possibly more extended blue loops.
\begin{figure}
\resizebox{\hsize}{!}{\includegraphics{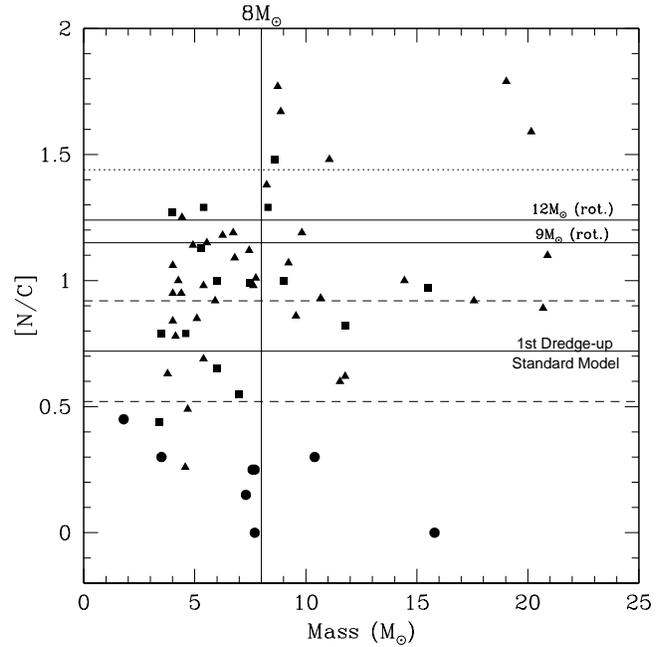}}
\caption{Plot of the [N/C] ratio vs. the mass of the stars. The circles are stars from Barbuy et al. (\cite{B96}), the triangles are stars from Luck \& Lambert (\cite{LL85}), and the squares are the stars from this work. The solid horizontal lines represent the predictions for [N/C] after the 1st dredge-up from Meynet \& Maeder (\cite{MeM00}) for stars without rotation and for stars of 9M$_{\odot}$ and 12M$_{\odot}$ with initial rotation of 300km s$^{-1}$, as indicated. The dashed lines represent the prediction without rotation $\pm$0.2 dex. The stars within these lines are considered to be fully mixed and in accordance with the non-rotating model. The dotted line represents the prediction of the 12M$_{\odot}$ rotating model +0.2 dex.}
\label{fig:ncmass}
\end{figure}
\begin{figure}
\resizebox{\hsize}{!}{\includegraphics{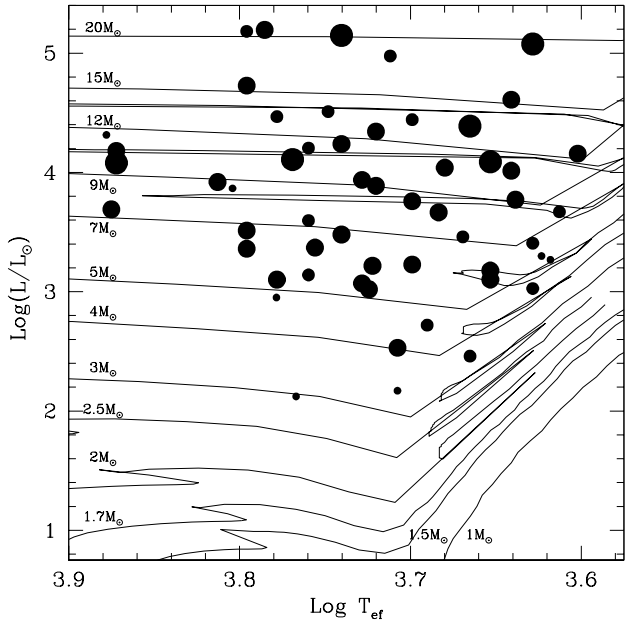}}
\caption{HR diagram with the stars from this work, from Luck \& Lambert (\cite{LL85}) and from Barbuy et al. (\cite{B96}) divided according to their abundances. The stars are represented by circles with sizes that are proportional to the [N/C] ratio. The smaller circles are the stars with [N/C] $<$ +0.52, the second smaller are stars with +0.52 $<$ [N/C] $<$ +0.92, the second larger the ones with +0.92 $<$ [N/C] $<$ +1.44, and the larger ones the stars with [N/C] $>$ 1.44. The evolutionary tracks are the ones by Schaller et al. (\cite{Sc92}).}
\label{fig:hr2}
\end{figure}
\begin{figure}
\resizebox{\hsize}{!}{\includegraphics{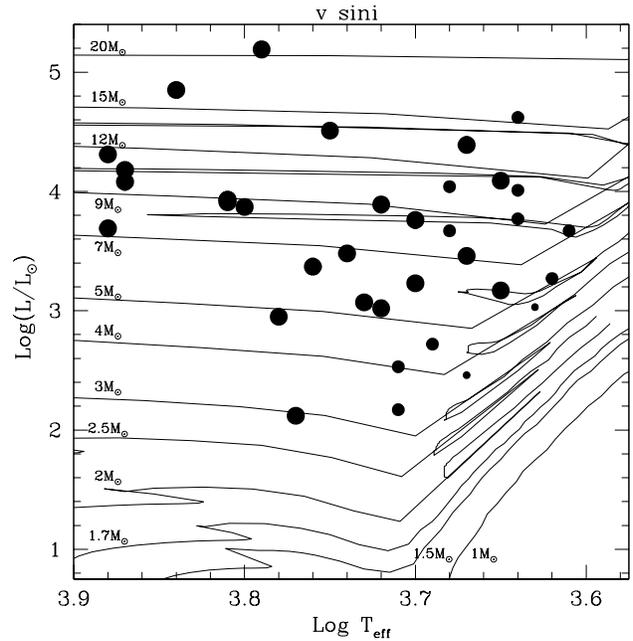}}
\caption{HR diagram with the stars from this work, from Luck \& Lambert (\cite{LL85}) and from Barbuy et al. (\cite{B96}) divided according to their $vsini$. The stars are represented as circles proportional to the $vsini$. The smaller circles are stars with $vsini$ $<$ 2.00km s$^{-1}$, the medium size circles are the stars with 2.00 km s$^{-1}$ $<$ $vsini$ $<$ 6.00 km s$^{-1}$, and the larger ones are stars with $vsini$ $>$ 6.00 km s$^{-1}$.}
\label{fig:hrvseni}
\end{figure}
\par We still have the stars  HD 45348, \object{HD 71181}, HD 76860, HD 80404, and HD 204867. All of them have [N/C] ratios that are only slightly higher than the limit of +0.92dex, for which we would consider [N/C] to be in agreement with the non-rotating models. Since the predictions of the Meynet \& Maeder (\cite{MeM00}) models are for stars with a limiting high rotation (v = 300 km s$^{-1}$), the abundances of these stars would be in agreement with a mixing induced by a lower rotation. Although these abundances seem to indicate more efficient mixing, we have to keep in mind that the uncertainties of the abundances do not allow a firm conclusion.
\par Figure \ref{fig:ncmass} is a plot of the [N/C] ratio vs. stellar mass with the stars of our sample, of the sample analyzed by Luck \& Lambert (\cite{LL85}) and of the sample by Barbuy et al. (\cite{B96}). Masses for the stars of Luck \& Lambert (\cite{LL85}) were estimated as in Barbuy et al. (\cite{B96}) with the mass-luminosity relation by Schaller et al. (\cite{Sc92}).  As for our sample we can see the existence of stars spanning a variety of mixing efficiences. The are stars only partially mixed, stars fully mixed and in agreement with the non-rotating model, stars fully mixed, but beyond what is predicted by the standard model, that can in principle be explained by the rotating models. We can also identify a fourth group of stars that seem to be mixed beyond what is  predicted by the rotating models.
\par In Fig. \ref{fig:ncmass} only stars that are more massive than about 8M$_{\odot}$ seem to be more mixed than expected for the rotating models. We can also note some 5M$_{\odot}$ stars that are mixed as predicted for the 12M$_{\odot}$ stars. These indicate that, although the rotation-induced mixing included in the models can produce [N/C] ratios compatible with the observations, there are details in the mass dependence of the results yet to be explored. Of course this should be seen with caution since accurately determining stellar masses is not an easy matter.
\par Figure \ref{fig:hr2} shows the distribution along the HR diagram of the same stars plotted in Fig. \ref{fig:ncmass}. This time, however, the stars are divided in four groups according to their abundances. We note that the different group of stars overlap, which is not difficult to understand since the blue loops occur in this region of the HR diagram. However we also note a group of fully mixed stars with masses around 5M$_{\odot}$ again in a region where blue loops are not expected. We cannot state whether this is due to an incorrect estimation of the masses or to an underestimated extent of the blue loops.
\par Figure \ref{fig:hrvseni} shows the distribution along the HR diagram only for the stars for which $vsini$ have been determined in the literature. The stars are again represented as circles, this time with sizes proportional to their $vsini$. Very interesting to note is that in Fig. \ref{fig:hrvseni} the stars with smaller $vsini$ tend to be concentrated towards the right side of the HR diagram. This is an expected result, since the $vsini$ is expected to decrease with increasing radii, and lower T$_{\rm{eff}}$, during the evolution of the star. The observed $vsini$ thus is not the initial $vsini$ that would drive the mixing in the main sequence.
\begin{figure*}
\centering
\includegraphics[width=17cm]{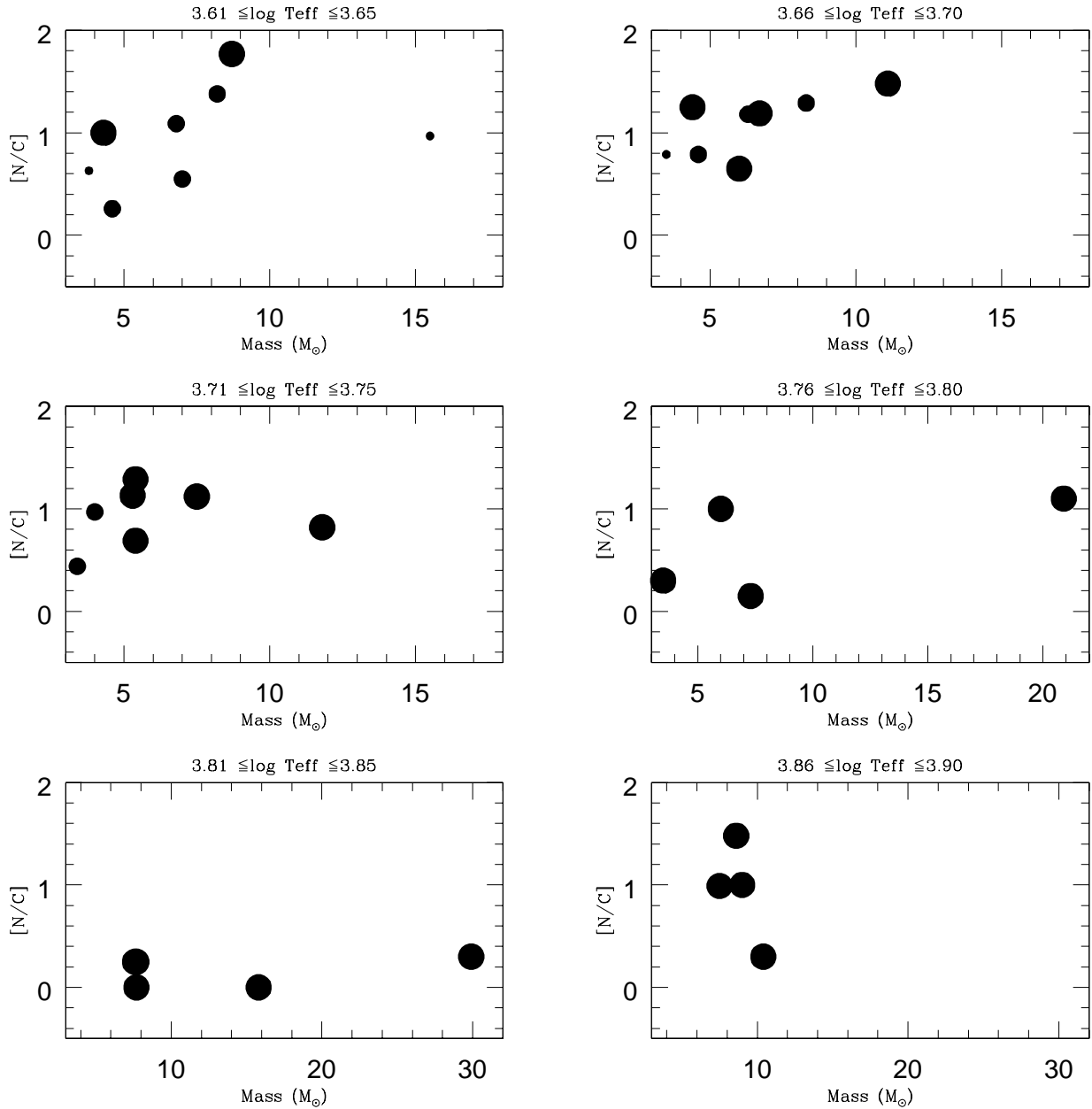}
\caption{[N/C] vs. stellar mass divided in intervals of temperature. The symbols are related to the $vsini$ in the same manner as in the last plot.}
\label{fig:bintef}
\end{figure*}
\begin{figure}
\resizebox{\hsize}{!}{\includegraphics{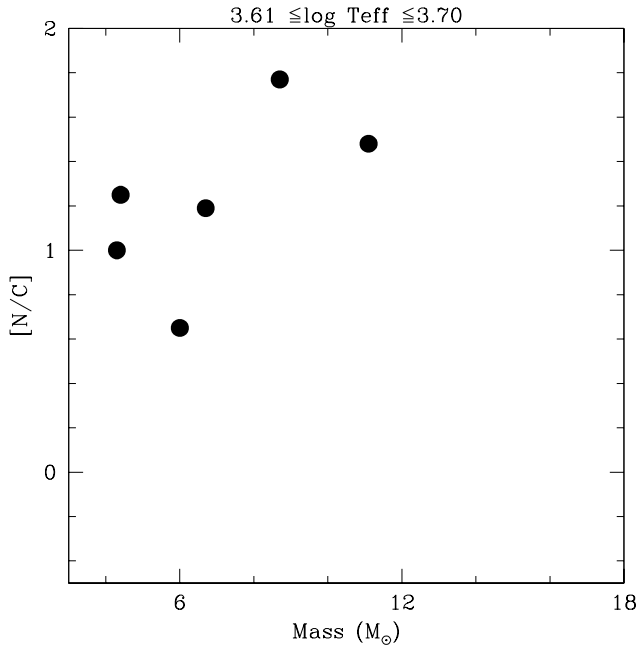}}
\caption{Plot of [N/C] vs. stellar mass in the interval between log T$_{\rm{eff}}$ = 3.61 and log T$_{\rm{eff}}$ = 3.70 where a clear correlation can be noted. In this figure only the stars with $vsini$ $>$ 6 km s$^{-1}$ are plotted. }
\label{fig:ncmasscor}
\end{figure}
\begin{figure}
\resizebox{\hsize}{!}{\includegraphics{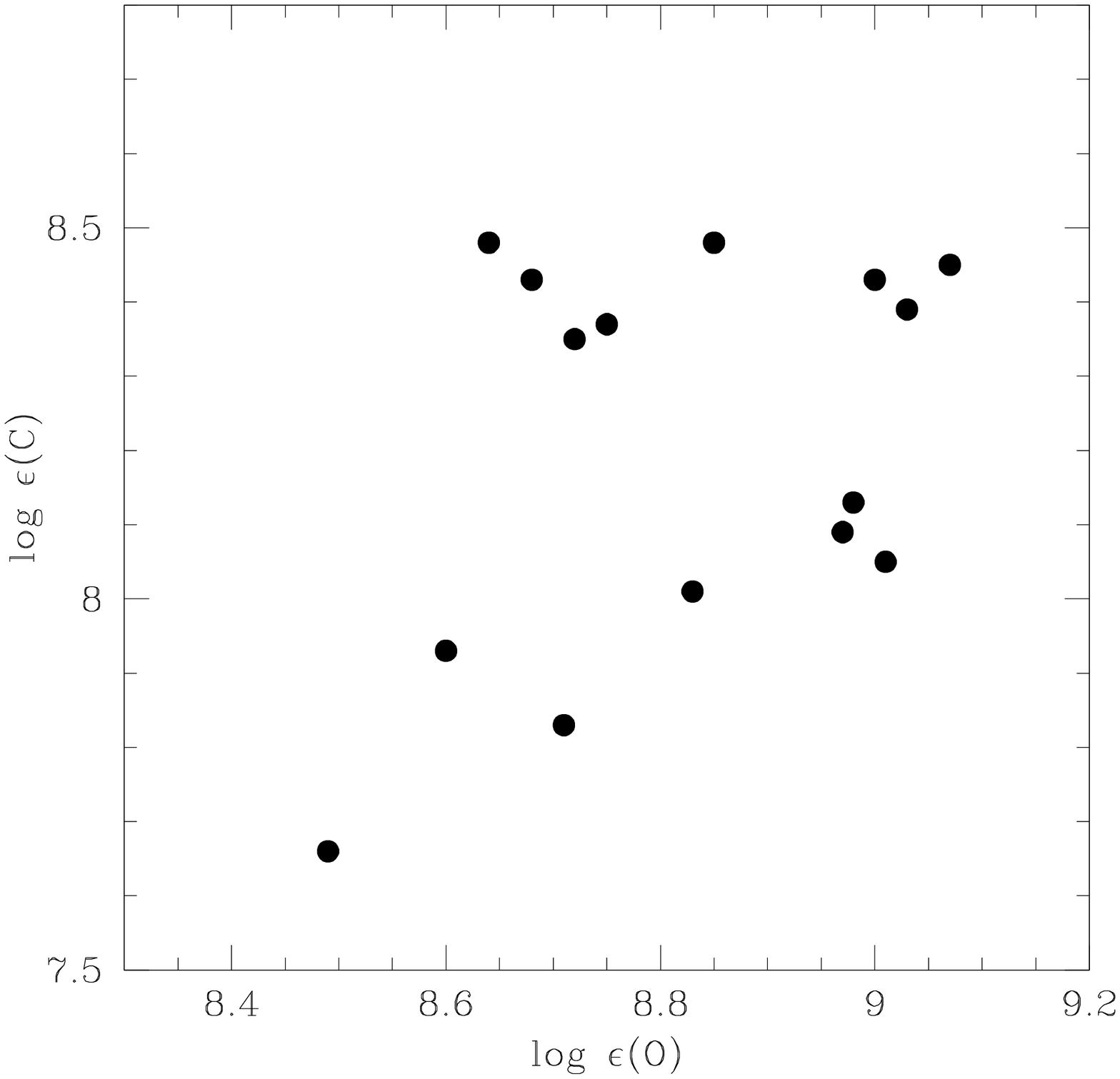}}
\caption{Plot of the oxygen abundance vs. the carbon abundance. There is no correlation between them.}
\label{fig:oxicar}
\end{figure}
\par The [N/C] values depend both on the stellar mass and on its initial $vsini$. In order to try to disentangle the various effects, we plotted [N/C] vs. mass in intervals of temperature representing the stars again as circles with sizes proportional to $vsini$ (Fig. \ref{fig:bintef}). Some hints of trends can readily be seen, but in none of the plots are the trends strong. However, when considering the interval between log T$_{\rm{eff}}$ = 3.61 and log T$_{\rm{eff}}$ = 3.70, the red supergiants, and only the stars with $vsini$ $>$ 6 km s$^{-1}$ an important correlation can be seen (Fig \ref{fig:ncmasscor}). This correlation indicates a large increase in [N/C] with increasing mass.
\par It is the first time that such a relation is obtained and it may represent a new and important constraint. One has to still keep in mind that the correlation is defined by only a small number of points and that there seems to be a large scatter. The scatter, however, is probably mostly due to the fact that a large number of the stars will reach the giant branch with $vsini$ $>$ 6 km s$^{-1}$ and not only to observational uncertainties. Further work in extending the sample and confirming the correlation and the scatter is still needed.

\par As a last point, we recall that Luck \& Lambert (\cite{LL85}) analyzed a sample of variable and non-variable cool supergiants and noted a correlation between the carbon and oxygen abundances of their sample stars. Based on this correlation, they argued that the dredge-up could possibly reach deeper regions where the action of the ON cycle would be important for the final abundances. In order to investigate this suggested correlation, we made a plot of carbon vs. oxygen for our sample (Fig. \ref{fig:oxicar}). There is no indication of correlation. Hence, there seems to be no need of any deep mixing to explain the abundances.

\section{Conclusions}

We carried out a detailed analysis of a sample of 19 evolved intermediate mass stars using high resolution spectra. We have determined atmospheric parameters, masses, and CNO abundances using spectral synthesis. Fifteen stars for which we were able to determine both carbon and nitrogen show signs of internal mixing. The mean [N/C] ratio found is [N/C] = +0.95dex.
\par Only five of these stars, HD 49068, HD 51043, HD 102839, HD 114792, and HD 225212, have abundances in agreement with the predictions of non-rotating models by Meynet \& Maeder (\cite{MeM00}), which predict [N/C] = +0.72dex. One other star, HD 38713, would be less mixed than this but agrees marginally with the predictions of non-rotating models by Schaller et al.\ (\cite{Sc92}), [N/C] = +0.60dex. 
\par All the other stars show signs of a more efficient mixing process, i.e., [N/C] ratios higher than expected. The rotating models by Meynet \& Maeder (\cite{MeM00}) seem to show better agreement with the observed abundances. There seems to be, however, somewhat poor agreement when we consider the run of the abundances with the stellar mass, a point that needs further investigation.
\par Five of our stars, HD 45348, HD 71181, HD 76860, HD 80404, and HD 204867, have [N/C] $\approx$ +1.0dex, less than what is predicted from rotating models with v = 300km s$^{-1}$. Thus, these stars might have lower rotation. Two other stars, HD 36673 and HD 66190, seem to have even higher [N/C] than predicted by these models. All this clearly indicates that the recent efforts to introduce rotation effects in the evolutionary models are producing results that are more compatible with observations.
\par The distribution of the stars in the HR diagram (Fig. \ref{fig:hr2}) seems to indicate  that the extension of the blue loops is underestimated at least in the 5M$_{\odot}$ track,  further consideration of the uncertainties affecting the masses and luminosities of the stars in this region is needed before a firm conclusion can be drawn.
\par A correlation between [N/C] and stellar mass in the interval between log T$_{\rm{eff}}$ = 3.61 and log T$_{\rm{eff}}$ = 3.70 has been indentified for the first time. This might represent an important constraint for the efforts at including rotation in the models. The correlation, however, is defined by a small number of points and should be investigated further.
\par We would like to emphasize that more observations are still needed. The CNO abundances, along with other elements that are affected by mixing processes such as Na and Li, must be determined in extended samples of A, F, G, K, and M evolved stars with low and high $vsini$. Only in this way will we be able to trace the mixing processes along the HR diagram and to better constrain the influence of rotation in the evolutionary models.

\begin{acknowledgements}

RS acknowledges a CAPES fellowship during the development of this work based on his MSc thesis. BB acknowledges grants from the Instituto do Mil\^enio - CNPq, 620053/2001-1 and the FAPESP. JRM acknowledges financial support from the Instituto do Mil\^enio - CNPq, 620053/2001-1 and the FAPERN Agency. The FEROS observations at the European Southern Observatory (ESO) were
carried out within the Observat\'orio Nacional ON/ESO and ON/IAG agreements, under Fapesp project n$^{\circ}$ 1998/10138-8.

\end{acknowledgements}



\Online
%

\begin{longtable}{ccccccc}
\caption{ \label{LE} FeI and FeII equivalent widths for the stars HD1219, HD36673
HD38713, HD44362, and HD45348} \\
\hline\hline
$\lambda$ &  Species & HD1219 &  HD36673 & HD38713 & HD44362  & HD45348 \\
\hline
\endfirsthead
\caption{continued.} \\
\hline\hline
$\lambda$ &  Species & HD1219 &  HD36673 & HD38713 & HD44362  & HD45348 \\
\hline
\endhead
\hline
\endfoot
5256.94  & FE2  &   --   &  98.7  &  53.1  & 106.0  &  65.8 \\   
5264.81  & FE2  &  71.0  & 192.3  &  86.5  & 192.1  &   --  \\   
5276.00  & FE2  &   --   & 331.4  & 203.9  &   --   &   --  \\
5284.11  & FE2  &  84.8  & 211.3  & 105.9  & 218.9  & 161.0 \\  
5320.040 & FE1  &   --   &   --   &  44.7  &   --   &   --  \\ 
5321.109 & FE1  &  78.6  &  10.2  &  67.2  &   --   &   --  \\
5325.56  & FE2  &  60.1  & 172.7  &  85.6  & 182.0  & 127.7 \\   
5337.73  & FE2  &   --   &   --   &   --   & 142.7  &   --  \\ 
5362.87  & FE2  & 158.3  & 259.4  & 166.6  & 267.5  &   --  \\   
5364.880 & FE1  & 162.1  &  91.9  & 151.9  & 197.4  &  86.0 \\   
5365.407 & FE1  &   --   &  30.9  & 113.0  & 145.7  &  23.7 \\   
5367.476 & FE1  & 170.7  & 114.0  & 159.3  & 206.4  & 110.5 \\   
5369.974 & FE1  & 205.9  & 123.2  & 179.4  & 241.4  & 127.1 \\   
5373.714 & FE1  & 100.9  &  23.6  &  91.5  & 109.1  &  31.4 \\   
5379.581 & FE1  & 111.7  &  25.9  & 102.1  & 108.6  &  15.0 \\   
5383.380 & FE1  & 205.4  & 137.8  & 197.9  & 246.6  & 142.0 \\   
5386.340 & FE1  &  79.4  &   1.8  &  60.8  &  53.8  &   6.6 \\   
5389.486 & FE1  &   --   &  50.9  & 120.4  & 156.7  &  56.8 \\   
5393.176 & FE1  & 202.3  &  82.8  & 183.3  &   --   &  80.8 \\   
5395.222 & FE1  &   --   &   --   &   --   &   --   &   --  \\ 
5397.623 & FE1  &   --   &   --   &   --   &   --   &   7.6 \\
5398.287 & FE1  & 114.9  &  34.4  & 105.4  & 130.0  &  43.7 \\   
5400.511 & FE1  &   --   &  74.5  & 189.7  &   --   &  57.9 \\ 
5412.791 & FE1  &  62.4  &   --   &  44.8  &   --   &   --  \\
5414.07  & FE2  &  50.5  & 121.9  &  65.8  & 138.3  &  88.8 \\   
5425.26  & FE2  &  65.7  & 165.5  &  76.7  & 167.5  & 121.3 \\   
5432.97  & FE2  &   --   & 142.6  &   --   &   --   &   --  \\  
5436.297 & FE1  &  77.7  &   --   &  66.5  &   --   &   --  \\ 
5473.168 & FE1  &   --   &   --   &  41.8  &   --   &   --  \\ 
5483.108 & FE1  &   --   &   7.9  &  72.1  &   --   &  15.7 \\ 
5491.845 & FE1  &  47.6  &   --   &  25.3  &   --   &   --  \\ 
5494.474 & FE1  &  73.5  &   --   &  55.3  &   --   &   --  \\ 
5506.791 & FE1  & 257.4  &  76.1  & 215.9  & 290.7  &  90.0 \\
5522.454 & FE1  &  83.6  &   --   &  69.3  &  54.5  &  13.6 \\   
5525.13  & FE2  &   --   &  66.5  &  39.4  &  72.9  &  47.6 \\   
5525.552 & FE1  &  99.7  &   --   &  83.1  &  81.6  &  20.6  \\
5531.985 & FE1  &   --   &   --   &  30.8  &   --   &   --   \\  
5534.848 & FE2  &   --   & 225.0  & 100.4  & 226.9  & 223.9  \\  
5539.291 & FE1  &  66.5  &   --   &  38.6  &   --   &   --   \\  
5543.944 & FE1  &  99.1  &   --   &  85.4  &  90.7  &  38.9  \\  
5546.514 & FE1  &   --   &  16.3  &  71.2  &  68.6  &   --   \\  
5560.207 & FE1  &   --   &  19.9  &  76.7  &  76.2  &   --   \\  
5577.013 & FE1  &   --   &   --   &   --   &   --   &   --  \\ 
5587.573 & FE1  &  81.5  &   --   &  64.0  &   --   &   --  \\
5635.824 & FE1  &   --   &   --   &   --   &   --   &   --  \\ 
5636.705 & FE1  &  66.5  &   --   &  45.5  &   --   &   --   \\  
5638.262 & FE1  & 125.9  &  37.2  & 108.5  & 137.5  &  27.5  \\
5641.436 & FE1  &   --   &   --   &   --   & 106.0  &  19.2  \\  
5646.697 & FE1  &  42.3  &   --   &   --   &   --   &   --   \\ 
5650.019 & FE1  &  72.1  &   --   &  57.5  &   --   &  11.0  \\
5652.319 & FE1  &  67.9  &   --   &  48.9  &  36.2  &   --   \\ 
5661.348 & FE1  &   --   &   --   &  49.3  &  36.4  &   --   \\ 
5680.240 & FE1  &   --   &   --   &   --   &   --   &   --  \\
5701.557 & FE1  & 157.9  &  24.1  & 128.6  &   --   &  19.0 \\ 
5705.473 & FE1  &  79.8  &   --   &  61.2  &  50.6  &   --  \\  
5731.761 & FE1  &  96.1  &  18.9  &  89.7  &  87.3  &   --  \\   
5738.240 & FE1  &  51.9  &   --   &  26.8  &   --   &   --  \\ 
5778.463 & FE1  &  82.1  &   --   &  54.5  &   --   &   --  \\ 
5784.666 & FE1  &  82.6  &   --   &  58.8  &   --   &   --  \\
5811.916 & FE1  &   --   &   --   &  26.9  &   --   &   --  \\ 
5814.805 & FE1  &  62.5  &   --   &  45.3  &  34.4  &   --  \\ 
5835.098 & FE1  &  51.8  &   --   &   --   &   --   &   --  \\
6012.212 & FE1  &   --   &   --   &  55.6  &   --   &   --  \\ 
6079.014 & FE1  &  79.1  &  14.5  &  73.1  &   --   &   --  \\ 
6084.10  & FE2  &  43.4  & 104.5  &  57.4  & 111.9  &  70.5 \\
6093.666 & FE1  &  66.7  &   --   &  53.7  &  41.8  &   5.4 \\   
6098.250 & FE1  &  53.4  &   --   &   --   &   --   &   5.1 \\ 
6113.33  & FE2  &  27.9  &  66.6  &  39.7  &   --   &  43.2 \\
6120.249 & FE1  &   --   &   --   &  33.5  &   --   &   --  \\ 
6129.70  & FE2  &   --   &  32.7  &   --   &   --   &  20.6 \\ 
6136.615 & FE1  &   --   &  81.1  & 186.6  & 234.3  &  65.3 \\
6137.002 & FE1  & 142.1  &   --   & 118.5  &   --   &  10.9 \\ 
6137.702 & FE1  & 250.2  &  72.6  & 207.6  & 266.3  &  67.9 \\ 
6141.03  & FE2  &  17.2  &  14.8  &  15.2  &   --   & 150.2 \\   
6149.25  & FE2  &  46.4  & 160.6  &  74.3  & 164.3  & 115.2 \\ 
6151.616 & FE1  &   --   &   --   &  98.2  &  87.1  &   --  \\ 
6157.733 & FE1  &   --   &  28.4  & 101.4  & 119.0  &  18.9 \\   
6159.382 & FE1  &  48.9  &   --   &  29.8  &   --   &   --  \\  
6165.363 & FE1  &  87.3  &   9.6  &  77.5  &  66.2  &   --  \\   
6170.500 & FE1  &   --   &  31.3  &   --   & 115.2  &  26.0 \\ 
6173.341 & FE1  & 137.7  &  20.6  & 118.1  & 125.0  &   --  \\   
6179.39  & FE2  &   --   &  41.6  &   --   &   --   &  25.9 \\ 
6187.987 & FE1  &  93.9  &   --   &  79.8  &  71.5  &   --  \\ 
6191.558 & FE1  &   --   &  88.9  & 183.1  & 227.8  &  56.5 \\
6213.428 & FE1  & 156.0  &   --   &   --   & 154.4  &   --   \\  
6226.730 & FE1  &  72.1  &   --   &  56.8  &   --   &   --  \\ 
6238.39  & FE2  &  63.9  &   --   &  85.2  & 172.7  & 130.4 \\
6239.95  & FE2  &   --   &   --   &   --   &   --   &  58.5 \\ 
6240.645 & FE1  &   --   &   --   &   --   &  84.2  &   --   \\  
6247.55  & FE2  &  53.8  &   --   &  89.7  & 219.2  &   --  \\
6248.90  & FE2  &   --   &   --   &   --   &   --   &   --  \\   
6271.283 & FE1  &  75.6  &   --   &  52.0  &   --   &   --  \\ 
6290.974 & FE1  &   --   &   --   &   --   &  87.9  &  14.8 \\
6293.933 & FE1  &   --   &   --   &   --   &   --   &   --  \\ 
6297.799 & FE1  & 149.3  &  13.9  &   --   & 142.8  &   --  \\   
6301.508 & FE1  & 167.8  &   --   & 163.0  &   --   &  63.7  \\  
6302.499 & FE1  & 134.8  &  29.2  &   --   &   --   &  19.3  \\  
6305.314 & FE2  &   --   &   --   &   --   &  38.4  &   --   \\  
6307.854 & FE1  &   --   &   --   &   --   &   --   &   --   \\ 
6315.314 & FE1  &   --   &   --   &   --   &   --   &   --  \\ 
6315.814 & FE1  &  93.3  &   9.2  &  79.5  &   --   &  15.2 \\ 
6322.694 & FE1  & 143.3  &  24.5  &   --   & 135.4  &   9.7  \\  
6331.95  & FE2  &   --   &  44.4  &   --   &   --   &  35.9  \\  
6336.823 & FE1  &   --   &  45.2  & 136.2  & 161.9  &  37.8  \\  
6369.46  & FE2  &  30.8  &  96.3  &  50.6  & 106.5  &  68.4 \\
6380.750 & FE1  & 103.1  &   --   &  82.3  &  75.1  &  19.0  \\  
6383.72  & FE2  &   --   &  74.1  &   --   &   --   &   --  \\ 
6385.45  & FE2  &   --   &  56.5  &   --   &   --   &  32.3 \\
6385.726 & FE1  &  39.2  &   --   &   --   &   --   &   --  \\ 
6392.538 & FE1  &  80.8  &   --   &   --   &   --   &   --  \\ 
6393.602 & FE1  &   --   &  83.3  & 180.3  & 233.0  &  60.1 \\
6400.000 & FE1  &   --   &  92.1  & 173.4  &   --   &  68.6 \\   
6411.658 & FE1  & 193.7  &  73.7  & 160.7  & 192.1  &  77.9 \\ 
6416.92  & FE2  &  57.1  & 169.9  &  68.4  & 156.9  & 118.8 \\ 
6419.956 & FE1  & 128.8  &  57.8  &   --   & 132.1  &  51.8  \\  
6421.360 & FE1  & 200.7  &  62.6  & 220.5  & 239.0  &  50.1  \\  
6430.856 & FE1  & 231.9  &  60.7  & 181.6  & 237.6  &  50.5  \\  
6432.68  & FE2  &  56.9  & 153.7  &  82.7  & 170.5  & 113.5 \\   
6442.94  & FE2  &   --   &  52.5  &   --   &   --   &  36.7  \\  
6446.40  & FE2  &   --   &  48.8  &   8.0  &   --   &  33.8  \\  
6456.39  & FE2  &  68.8  &   --   & 105.7  & 254.5  &   --   \\ 
\hline
\end{longtable}
\begin{longtable}{ccccccc}
\caption{ \label{LE2} FeI and FeII equivalent widths for the stars HD49068, HD49396, 
HD51043, HD66190, and HD71181} \\
\hline\hline
$\lambda$ &  Species & HD49068 &  HD49396 & HD51043 & HD66190  & HD71181 \\
\hline
\endfirsthead
\caption{continued.} \\
\hline\hline
$\lambda$ &  Species & HD49068 &  HD49396 & HD51043 & HD66190  & HD71181 \\
\hline
\endhead
\hline
\endfoot
5256.94  & FE2  &   --   & 128.0  &  76.3  &  86.1 &  79.5 \\  
5264.81  & FE2  &  78.0  & 225.9  & 123.7  & 124.1 & 113.9 \\  
5276.00  & FE2  &   --   &   --   &   --   &       & 257.2 \\
5284.11  & FE2  & 103.8  & 228.7  &   --   & 149.3 & 154.1 \\  
5320.040 & FE1  &  75.5  &   --   &   --   &   --  &   --  \\ 
5321.109 & FE1  &  92.3  &  98.6  &  95.8  & 107.6 &   --  \\
5325.56  & FE2  &  80.9  &   --   & 120.7  & 129.0 & 123.1 \\  
5337.73  & FE2  &   --   &   --   &   --   &  99.3 &  --   \\ 
5362.87  & FE2  & 178.4  & 303.8  & 219.7  & 230.6 & 211.9 \\  
5364.880 & FE1  & 168.3  & 300.1  & 192.4  & 229.6 & 183.0 \\  
5365.407 & FE1  & 141.1  & 188.1  & 158.1  & 183.7 & 149.8 \\  
5367.476 & FE1  & 185.8  & 307.9  & 222.4  & 248.0 & 190.6 \\  
5369.974 & FE1  & 227.6  & 314.9  & 231.0  & 304.8 & 215.3 \\  
5373.714 & FE1  & 115.3  & 134.0  & 123.7  & 143.0 & 112.4 \\  
5379.581 & FE1  & 137.5  & 155.8  & 140.8  & 163.3 & 123.0 \\  
5383.380 & FE1  & 241.5  & 331.9  & 260.4  & 299.5 & 210.5 \\  
5386.340 & FE1  &  84.6  &  87.2  &  81.8  & 102.8 &  69.4 \\  
5389.486 & FE1  & 145.9  & 198.3  & 152.3  & 175.8 & 140.0 \\  
5393.176 & FE1  & 230.1  & 339.2  &   --   & 294.3 & 212.1 \\  
5395.222 & FE1  &   --   &   --   &  52.0  &   --  &  45.9 \\ 
5397.623 & FE1  &   --   &   --   &  90.3  &   --  &  72.8 \\
5398.287 & FE1  & 127.8  & 169.7  & 140.1  & 169.0 & 121.1 \\  
5400.511 & FE1  & 249.5  & 248.8  & 234.2  & 273.7 & 218.0 \\ 
5412.791 & FE1  &  61.8  &   --   &   --   &  86.1 &  52.6 \\
5414.07  & FE2  &  51.5  & 146.9  &  89.2  &  88.0 &  80.7 \\  
5425.26  & FE2  &  68.7  & 186.9  & 116.9  & 106.4 &  --   \\  
5432.97  & FE2  &   --   &   --   &   --   & 151.8 &  --   \\  
5436.297 & FE1  &  78.4  &   --   &   --   &   --  &  74.1 \\ 
5473.168 & FE1  &  67.5  &   --   &   --   &   --  &   --  \\ 
5483.108 & FE1  &   --   &   --   &   --   &   --  &   --  \\ 
5491.845 & FE1  &   --   &   --   &   --   &  57.7 &  47.9 \\ 
5494.474 & FE1  &  79.1  &  79.8  &  76.2  &  98.9 &  83.3 \\ 
5506.791 & FE1  & 348.8  & 442.0  & 366.9  & 406.6 & 295.7 \\
5522.454 & FE1  &  91.8  & 115.4  & 103.0  & 127.4 &  91.6 \\  
5525.13  & FE2  &   --   &  97.9  &  67.6  &   --  &   --  \\  
5525.552 & FE1  & 115.6  & 144.7  & 120.4  & 139.7 & 110.0  \\
5531.985 & FE1  &   --   &   --   &   --   &  73.5 &  47.1  \\ 
5534.848 & FE2  & 111.4  & 341.0  & 155.2  & 182.4 & 156.9  \\ 
5539.291 & FE1  &  75.7  &   --   &  74.3  &  92.6 &  60.4  \\ 
5543.944 & FE1  & 112.3  & 143.4  & 119.9  & 143.5 & 115.7  \\ 
5546.514 & FE1  &   --   & 117.4  & 108.4  & 139.3 & 106.3  \\ 
5560.207 & FE1  &   --   & 111.8  & 102.4  & 118.0 &  91.7  \\ 
5577.013 & FE1  &   --   &   --   &   --   &  --   &  21.7 \\ 
5587.573 & FE1  &  89.5  &   --   &  82.8  &  --   &  74.4 \\
5635.824 & FE1  &  82.3  &  83.4  &  84.4  & 107.0 &  78.3 \\ 
5636.705 & FE1  &  77.7  &   --   &  66.6  &  96.6 &  64.8  \\ 
5638.262 & FE1  & 137.1  & 173.9  & 142.2  & 175.8 & 142.8  \\
5641.436 & FE1  &   --   & 139.1  & 131.6  & 161.5 & 128.7  \\ 
5646.697 & FE1  &  38.3  &   --   &   --   &  41.6 &   --   \\ 
5650.019 & FE1  &  76.7  &   --   &  80.9  &  90.9 &  70.5  \\
5652.319 & FE1  &  70.1  &   --   &  65.7  &  86.9 &  62.1  \\ 
5661.348 & FE1  &   --   & 64.0   &  73.3  &  96.9 &  62.0  \\ 
5680.240 & FE1  &   --   &   --   &   --   &  66.6 &  39.5 \\
5701.557 & FE1  & 185.9  & 215.2  & 186.1  & 232.3 & 181.1 \\ 
5705.473 & FE1  &  83.5  &  86.6  &  80.5  & 104.7 &  85.2 \\  
5731.761 & FE1  & 120.2  & 136.3  & 121.9  & 144.2 & 109.5 \\  
5738.240 & FE1  &  55.0  &   --   &  43.6  &  --   &  36.1 \\ 
5778.463 & FE1  & 100.6  &  72.0  &   --   & 116.3 &  79.1 \\ 
5784.666 & FE1  &  93.4  &   --   &  82.3  & 108.2 &  75.6 \\
5811.916 & FE1  &  44.4  &   --   &   --   &  51.4 &  35.0 \\ 
5814.805 & FE1  &  71.0  &  59.7  &   --   &  78.8 &  56.6 \\ 
5835.098 & FE1  &   --   &   --   &   --   &  --   &  40.2 \\
6012.212 & FE1  &  98.3  &   --   &   --   & 124.4 &  84.8 \\ 
6079.014 & FE1  &  94.0  &  99.0  &  95.4  & 120.7 &  87.5 \\ 
6084.10  & FE2  &  54.4  & 129.7  &  84.5  &  89.5 &  82.8 \\
6093.666 & FE1  &  76.5  &  64.2  &  71.6  &  88.1 &  67.4 \\  
6098.250 & FE1  &  62.7  &   --   &   --   &  75.5 &  48.1 \\ 
6113.33  & FE2  &  47.0  &   --   &   --   &  79.0 &  59.4 \\
6120.249 & FE1  &   --   &   --   &   --   &  99.4 &  51.1 \\ 
6129.70  & FE2  &   --   &   --   &   --   &  --   &  --   \\ 
6136.615 & FE1  & 258.5  &   --   & 257.0  &  --   & 246.3 \\
6137.002 & FE1  & 171.6  &   --   & 156.5  &  --   & 155.4 \\ 
6137.702 & FE1  & 307.5  & 340.1  & 304.7  & 367.3 & 268.7 \\ 
6141.03  & FE2  &  21.5  &   --   &   --   &   --  &   --  \\  
6149.25  & FE2  &  64.8  & 178.4  & 109.2  & 106.5 &  97.7 \\ 
6151.616 & FE1  & 148.7  & 152.1  &   --   &   --  & 128.5 \\ 
6157.733 & FE1  & 150.7  & 172.8  & 151.1  & 189.8 & 142.3 \\  
6159.382 & FE1  &  52.2  &   --   &  46.5  &  61.5 &  36.2 \\  
6165.363 & FE1  &  99.6  & 109.7  & 101.2  & 127.5 &  94.4 \\  
6170.500 & FE1  &   --   & 171.3  & 160.7  &  --   & 157.6 \\ 
6173.341 & FE1  & 185.7  & 218.5  & 161.9  & 250.6 & 171.5 \\  
6179.39  & FE2  &   --   &   --   &   --   &    -- &   --  \\ 
6187.987 & FE1  & 108.3  & 117.8  &   --   & 130.0 &  99.0 \\ 
6191.558 & FE1  &   --   &   --   &   --   &  --   & 259.9 \\
6213.428 & FE1  & 196.6  & 222.9  & 189.8  & 239.9 & 191.2  \\ 
6226.730 & FE1  &   --   &  71.2  &   --   &  94.8 &  73.5 \\ 
6238.39  & FE2  &  89.0  & 195.7  &   --   & 128.2 & 117.7 \\
6239.95  & FE2  &   --   &   --   &   --   &   --  &  --   \\ 
6240.645 & FE1  & 145.1  & 148.3  &   --   & 185.6 & 137.3  \\ 
6247.55  & FE2  &  74.3  & 239.6  &   --   &  --   & 129.1 \\
6248.90  & FE2  &   --   &   --   &   --   &   --  &   --  \\  
6271.283 & FE1  &  92.2  &   --   &  77.1  & 105.7 &  78.8 \\ 
6290.974 & FE1  &   --   & 131.2  &   --   & 147.2 &   --  \\
6293.933 & FE1  &   --   &   --   &   --   &   --  &   --  \\ 
6297.799 & FE1  & 221.1  & 201.8  & 185.5  & 269.8 & 183.6 \\  
6301.508 & FE1  & 223.3  &   --   & 220.9  & 274.8 & 185.6  \\ 
6302.499 & FE1  & 150.1  & 191.2  & 155.7  & 223.1 & 157.5  \\ 
6305.314 & FE2  &   --   &   --   &   --   &   --  &  --   \\ 
6307.854 & FE1  &   --   &   --   &   --   &   --  &  --   \\ 
6315.314 & FE1  & 152.2  &   --   &   --   & 174.0 &  --   \\ 
6315.814 & FE1  & 111.0  & 129.7  &  99.3  & 127.3 &  93.9 \\ 
6322.694 & FE1  & 191.0  & 200.6  & 170.9  & 265.6 & 165.7  \\ 
6331.95  & FE2  &   --   &   --   &   --   &   --  &  --    \\ 
6336.823 & FE1  & 195.5  & 227.4  & 194.8  & 232.4 & 183.1  \\ 
6369.46  & FE2  &  44.4  & 126.3  &  72.3  &  75.8 &  73.5 \\
6380.750 & FE1  & 120.8  & 128.5  & 119.2  & 147.3 & 115.1  \\ 
6383.72  & FE2  &   --   &   --   &   --   &   --  &  --   \\ 
6385.45  & FE2  &   --   &   --   &   --   &   --  &  --   \\
6385.726 & FE1  &  37.6  &   --   &   --   &   --  &  --   \\ 
6392.538 & FE1  &  97.8  &  65.1  &   --   & 119.6 &  76.0 \\ 
6393.602 & FE1  & 267.2  &   --   & 262.0  & 337.6 & 256.3 \\
6400.000 & FE1  & 220.9  &   --   &   --   &   --  &   --  \\  
6411.658 & FE1  & 215.2  & 284.6  & 216.2  & 281.4 & 201.2 \\ 
6416.92  & FE2  &  67.8  &   --   &  91.9  & 101.2 &  94.3 \\ 
6419.956 & FE1  & 146.1  & 173.5  & 146.5  & 182.5 & 147.4  \\ 
6421.360 & FE1  & 313.4  & 325.5  &   --   & 333.5 & 248.1  \\ 
6430.856 & FE1  & 291.2  & 343.0  & 279.1  & 351.0 & 248.9  \\ 
6432.68  & FE2  &  69.3  & 182.3  & 114.9  & 117.3 & 113.8 \\  
6442.94  & FE2  &  11.7  &  30.1  &  17.0  &   --  &  --    \\ 
6446.40  & FE2  &   7.4  &   --   &   --   &   --  &  --    \\ 
6456.39  & FE2  &  98.4  & 272.2  &   --   & 153.8 & 151.0  \\ 
\hline
\end{longtable}

\begin{longtable}{ccccccc}
\caption{ \label{LE3} FeI and FeII equivalent widths for the stars HD76860, HD80404, 
HD90289, HD102839, and HD114792} \\
\hline\hline
$\lambda$ &  Species & HD71860 &  HD80404 & HD90289 & HD102839  & HD114792 \\
\hline
\endfirsthead
\caption{continued.} \\
\hline\hline
$\lambda$ &  Species & HD76860 &  HD80404 & HD90289 & HD102839  & HD114792 \\
\hline
\endhead
\hline
\endfoot

5256.94  &  FE2 &   77.7 &   --   &   --   &  109.6 &  126.2  \\ 
5264.81  &  FE2 &   90.4 &  132.6 &   49.3 &  131.1 &   --    \\ 
5276.00  &  FE2 &   --   &   --   &  206.6 &   --   &   --    \\
5284.11  &  FE2 &  113.5 &  148.7 &   64.3 &   --   &  241.2  \\ 
5320.040 &  FE1 &   --   &   --   &   77.5 &   --   &   --    \\ 
5321.109 &  FE1 &  124.1 &   --   &   87.0 &   --   &   --    \\
5325.56  &  FE2 &  101.7 &  117.4 &   54.2 &  173.0 &  198.2  \\ 
5337.73  &  FE2 &   77.0 &   --   &   --   &   --   &  165.7  \\ 
5362.87  &  FE2 &  213.9 &  181.2 &   --   &  268.1 &  296.9  \\ 
5364.880 &  FE1 &  208.5 &   71.4 &   --   &  242.1 &  221.2  \\ 
5365.407 &  FE1 &  186.9 &   23.0 &   --   &  212.3 &  162.4  \\ 
5367.476 &  FE1 &  209.0 &  103.5 &   --   &  251.8 &  224.0  \\ 
5369.974 &  FE1 &  269.9 &  105.9 &  171.4 &  267.6 &  252.4  \\  
5373.714 &  FE1 &  148.3 &   19.8 &  100.9 &  153.5 &  108.3  \\  
5379.581 &  FE1 &  166.0 &   --   &  109.2 &  171.0 &   --    \\ 
5383.380 &  FE1 &  282.9 &  120.7 &  201.9 &  281.0 &  252.0  \\  
5386.340 &  FE1 &  102.2 &   --   &   68.5 &  102.1 &   --    \\  
5389.486 &  FE1 &  181.2 &   36.1 &  133.9 &  201.0 &  159.9  \\ 
5393.176 &  FE1 &  279.5 &   --   &  218.6 &  295.0 &  266.1  \\ 
5395.222 &  FE1 &   78.7 &   --   &   56.5 &   --   &   31.9  \\ 
5397.623 &  FE1 &   --   &   --   &   --   &   --   &   --    \\
5398.287 &  FE1 &  163.1 &   26.2 &  104.1 &  171.2 &  135.6  \\  
5400.511 &  FE1 &   --   &   52.6 &   --   &   --   &  215.6  \\ 
5412.791 &  FE1 &   91.7 &   --   &   --   &   76.8 &   --    \\
5414.07  &  FE2 &   --   &   79.6 &   --   &  102.7 &  144.5  \\ 
5425.26  &  FE2 &   76.9 &  109.6 &   40.0 &  131.0 &  189.2  \\ 
5432.97  &  FE2 &  168.5 &   89.2 &   --   &  170.5 &  174.6  \\ 
5436.297 &  FE1 &   --   &   --   &   77.1 &   --   &   --    \\ 
5473.168 &  FE1 &   80.2 &   --   &   --   &   --   &   --    \\ 
5483.108 &  FE1 &   --   &   --   &   81.3 &   --   &   --    \\ 
5491.845 &  FE1 &   --   &   --   &   40.2 &   67.5 &   --    \\ 
5494.474 &  FE1 &   96.8 &   --   &   78.1 &   --   &   --    \\ 
5506.791 &  FE1 &   --   &   53.7 &  396.1 &  477.2 &  306.4  \\
5522.454 &  FE1 &  125.1 &   --   &   84.0 &  134.6 &   80.6  \\ 
5525.13  &  FE2 &   --   &   44.4 &   --   &   --   &   99.5  \\ 
5525.552 &  FE1 &  144.1 &   11.8 &  103.2 &  170.1 &  105.2   \\
5531.985 &  FE1 &   --   &   --   &   --   &   --   &   32.4   \\
5534.848 &  FE2 &   --   &  189.4 &   --   &  236.4 &  259.8   \\
5539.291 &  FE1 &  111.0 &   --   &   71.0 &   96.6 &   40.9   \\ 
5543.944 &  FE1 &  146.3 &   21.6 &  100.3 &  163.7 &  115.9   \\ 
5546.514 &  FE1 &  146.3 &   --   &   90.4 &   --   &   97.4   \\
5560.207 &  FE1 &   --   &   --   &   76.4 &  127.3 &   88.5   \\
5577.013 &  FE1 &   --   &   --   &   23.3 &   --   &   --    \\ 
5587.573 &  FE1 &   96.2 &   --   &   71.8 &   --   &   --    \\
5635.824 &  FE1 &  108.7 &   --   &   76.9 &   --   &   --    \\ 
5636.705 &  FE1 &  107.5 &   --   &   69.4 &   99.4 &   49.0   \\
5638.262 &  FE1 &  180.2 &   24.9 &  123.8 &  193.5 &  152.2   \\
5641.436 &  FE1 &  169.8 &   --   &  117.7 &  168.3 &  120.3   \\
5646.697 &  FE1 &   54.9 &   --   &   36.8 &   33.1 &   --     \\
5650.019 &  FE1 &   99.1 &   --   &   66.1 &   93.4 &   --     \\
5652.319 &  FE1 &   96.6 &   --   &   61.2 &   85.6 &   45.0   \\
5661.348 &  FE1 &   --   &   --   &   --   &   85.6 &   --     \\
5680.240 &  FE1 &   --   &   --   &   47.7 &   --   &   --    \\
5701.557 &  FE1 &   --   &   --   &  179.6 &  267.3 &  174.3  \\ 
5705.473 &  FE1 &  115.1 &   --   &   74.7 &  117.2 &   72.3  \\ 
5731.761 &  FE1 &  153.0 &   --   &   99.8 &  151.3 &  105.4  \\ 
5738.240 &  FE1 &   66.1 &   --   &   58.6 &   59.4 &   --    \\ 
5778.463 &  FE1 &  145.5 &   --   &  100.0 &  126.2 &   49.6  \\ 
5784.666 &  FE1 &  128.2 &   --   &   89.6 &  111.1 &   --    \\
5811.916 &  FE1 &   55.2 &   --   &   38.5 &   51.9 &   --    \\ 
5814.805 &  FE1 &   90.7 &   --   &   67.1 &   81.5 &   41.6  \\ 
5835.098 &  FE1 &   --   &   --   &   45.4 &   64.8 &   --    \\
6012.212 &  FE1 &  146.8 &   --   &  102.0 &  132.2 &   --    \\ 
6079.014 &  FE1 &  106.5 &   13.6 &   75.4 &  125.0 &   78.5  \\ 
6084.10  &  FE2 &   69.2 &   67.6 &   --   &  113.7 &  132.5  \\
6093.666 &  FE1 &   89.5 &   --   &   62.2 &   94.2 &   57.6  \\ 
6098.250 &  FE1 &   88.0 &   --   &   57.0 &   --   &   43.7  \\ 
6113.33  &  FE2 &   --   &   42.3 &   --   &   95.1 &   98.0  \\
6120.249 &  FE1 &   --   &   --   &  103.2 &   --   &   --    \\ 
6129.70  &  FE2 &   --   &   20.9 &   --   &   --   &   --    \\ 
6136.615 &  FE1 &   --   &   53.1 &  251.2 &   --   &   --    \\
6137.002 &  FE1 &   --   &   --   &  159.2 &   --   &   --    \\ 
6137.702 &  FE1 &  464.8 &   58.6 &  281.1 &  384.7 &  271.0  \\ 
6141.03  &  FE2 &   --   &    7.1 &    8.6 &   --   &   --    \\ 
6149.25  &  FE2 &   82.9 &  109.6 &   --   &  135.9 &  180.4  \\ 
6151.616 &  FE1 &   --   &   --   &  142.1 &  205.9 &  111.3  \\ 
6157.733 &  FE1 &  230.7 &   --   &  130.4 &  204.9 &  140.1  \\ 
6159.382 &  FE1 &   70.1 &   --   &   53.3 &   55.2 &   19.5  \\ 
6165.363 &  FE1 &  132.6 &    4.8 &   90.6 &  135.8 &   82.7  \\ 
6170.500 &  FE1 &   --   &   --   &   --   &  234.7 &  142.9  \\ 
6173.341 &  FE1 &  289.1 &   --   &  164.6 &  262.6 &  166.4  \\ 
6179.39  &  FE2 &   --   &   27.4 &   --   &   --   &   --    \\ 
6187.987 &  FE1 &  141.1 &   --   &  101.5 &  142.1 &   86.6  \\ 
6191.558 &  FE1 &   --   &   54.2 &   --   &   --   &   --    \\
6213.428 &  FE1 &   --   &   27.0 &  179.8 &  274.7 &  191.2   \\
6226.730 &  FE1 &   98.9 &   --   &   80.8 &  106.3 &   --    \\ 
6238.39  &  FE2 &   --   &   --   &   --   &   --   &  191.6  \\
6239.95  &  FE2 &   --   &   50.0 &   --   &   --   &  101.0  \\ 
6240.645 &  FE1 &  209.6 &   --   &  139.5 &  216.0 &  110.9   \\
6247.55  &  FE2 &   83.4 &   --   &   49.0 &  170.0 &  237.3  \\
6248.90  &  FE2 &   --   &   35.5 &   --   &   --   &   27.4  \\ 
6271.283 &  FE1 &  120.8 &   --   &   82.2 &  114.9 &   --    \\ 
6290.974 &  FE1 &  145.4 &   18.0 &   --   &  163.8 &   --    \\
6293.933 &  FE1 &   --   &   --   &   34.7 &   --   &   --    \\ 
6297.799 &  FE1 &  292.3 &   --   &  184.3 &  274.9 &  175.9  \\ 
6301.508 &  FE1 &  246.6 &   41.5 &  156.0 &  263.3 &  213.6   \\
6302.499 &  FE1 &  198.2 &   25.1 &  133.5 &  221.0 &   --     \\
6305.314 &  FE2 &   --   &   --   &   --   &   --   &   --     \\
6307.854 &  FE1 &   --   &   --   &   33.3 &   --   &   --     \\
6315.314 &  FE1 &  186.3 &   --   &   --   &  179.4 &   --    \\ 
6315.814 &  FE1 &  131.4 &   --   &   84.9 &  129.6 &   --    \\ 
6322.694 &  FE1 &  250.8 &   --   &  158.8 &  249.6 &  155.8   \\
6331.95  &  FE2 &   --   &   36.9 &   --   &   --   &   --     \\
6336.823 &  FE1 &   --   &   30.8 &  152.8 &  256.5 &  190.9   \\
6369.46  &  FE2 &   63.8 &   65.4 &   28.8 &  102.4 &  124.7  \\
6380.750 &  FE1 &  164.7 &   --   &   --   &  159.3 &   92.4   \\
6383.72  &  FE2 &   --   &   58.9 &   --   &   --   &   46.3  \\ 
6385.45  &  FE2 &   --   &   --   &   --   &   --   &   --    \\
6385.726 &  FE1 &   55.8 &   --   &   --   &   43.0 &   --    \\ 
6392.538 &  FE1 &   --   &   --   &   92.6 &  127.3 &   --    \\ 
6393.602 &  FE1 &   --   &   46.7 &  249.3 &   --   &  268.4  \\
6400.000 &  FE1 &   --   &   60.2 &  183.6 &   --   &   --    \\ 
6411.658 &  FE1 &  268.6 &   53.9 &  174.6 &  273.2 &  217.2  \\ 
6416.92  &  FE2 &   81.0 &  105.4 &   --   &  125.1 &  165.2  \\ 
6419.956 &  FE1 &  182.2 &   46.2 &  110.9 &  207.2 &  153.3   \\
6421.360 &  FE1 &  359.1 &   40.7 &  286.4 &  359.9 &   --     \\
6430.856 &  FE1 &  418.1 &   43.2 &  306.0 &  385.2 &  254.1   \\
6432.68  &  FE2 &   80.1 &   97.4 &   54.8 &  148.2 &  189.3  \\  
6442.94  &  FE2 &   --   &   34.1 &   --   &   --   &   26.9   \\
6446.40  &  FE2 &   --   &   31.0 &   --   &   --   &   --     \\
6456.39  &  FE2 &  113.7 &  177.0 &   50.1 &  199.4 &  279.2   \\
\hline
\end{longtable}
\begin{longtable}{cccccc}
\caption{ \label{LE4} FeI and FeII equivalent widths for the stars HD159633, HD192876, 
HD204867, and HD225212} \\
\hline\hline
$\lambda$ &  Species & HD159633 &  HD192876 & HD204860 & HD225212 \\
\hline
\endfirsthead
\caption{continued.} \\
\hline\hline
$\lambda$ &  Species & HD159633 &  HD192876 & HD204860 & HD225212 \\
\hline
\endhead
\hline
\endfoot
5256.94  &  FE2 &  129.4 &   97.4 &  120.0 &   86.4 \\  
5264.81  &  FE2 &  216.2 &  168.0 &  220.1 &   81.4 \\  
5276.00  &  FE2 &   --   &   --   &   --   &   --   \\
5284.11  &  FE2 &  214.1 &  196.4 &  237.5 &   --   \\  
5320.040 &  FE1 &   --   &   --   &   --   &   --   \\ 
5321.109 &  FE1 &   86.8 &   86.2 &   --   &  147.2 \\
5325.56  &  FE2 &   --   &  159.7 &  201.1 &  117.8 \\  
5337.73  &  FE2 &   --   &   --   &  165.6 &   85.8 \\ 
5362.87  &  FE2 &  303.1 &  259.7 &  291.3 &  243.3 \\  
5364.880 &  FE1 &  251.9 &  205.1 &  214.2 &  226.2 \\  
5365.407 &  FE1 &  197.7 &  165.3 &  153.2 &  204.6 \\  
5367.476 &  FE1 &  258.7 &  211.2 &  225.2 &  222.7 \\  
5369.974 &  FE1 &  270.4 &  230.5 &  253.2 &  243.8 \\  
5373.714 &  FE1 &  155.1 &  124.7 &  108.4 &  165.2 \\  
5379.581 &  FE1 &  168.0 &  134.9 &  112.1 &  176.3 \\  
5383.380 &  FE1 &  287.1 &   --   &  247.5 &  271.6 \\  
5386.340 &  FE1 &   99.0 &   84.1 &   64.0 &  113.1 \\  
5389.486 &  FE1 &  199.4 &  171.0 &  165.7 &  187.8 \\  
5393.176 &  FE1 &  313.8 &  254.8 &  266.6 &  278.9 \\  
5395.222 &  FE1 &   --   &   --   &   --   &   97.4 \\ 
5397.623 &  FE1 &  519.5 &   60.3 &   --   &   --   \\
5398.287 &  FE1 &  181.3 &  151.8 &  141.5 &  174.9 \\  
5400.511 &  FE1 &  258.9 &  235.0 &  204.8 &   --   \\ 
5412.791 &  FE1 &   --   &   --   &   --   &   --   \\
5414.07  &  FE2 &  145.1 &  113.5 &  149.4 &   70.9 \\  
5425.26  &  FE2 &  178.5 &  140.8 &  183.9 &   --   \\  
5432.97  &  FE2 &   --   &  158.6 &  176.1 &   --   \\  
5436.297 &  FE1 &   --   &   82.6 &   --   &   --   \\ 
5473.168 &  FE1 &   --   &   --   &   --   &   98.2 \\ 
5483.108 &  FE1 &   --   &   --   &   --   &   --   \\ 
5491.845 &  FE1 &   --   &   32.2 &   --   &   --   \\ 
5494.474 &  FE1 &   --   &   74.2 &   46.9 &   --   \\ 
5506.791 &  FE1 &  455.3 &  350.0 &  311.2 &   --   \\
5522.454 &  FE1 &  117.2 &  100.8 &   77.7 &  138.5 \\  
5525.13  &  FE2 &   94.5 &   73.6 &   85.8 &   80.3 \\  
5525.552 &  FE1 &  141.9 &   --   &  103.9 &  159.6  \\
5531.985 &  FE1 &   --   &   --   &   --   &   --    \\ 
5534.848 &  FE2 &  267.5 &  213.9 &  255.0 &   --    \\ 
5539.291 &  FE1 &   67.2 &   58.1 &   33.0 &  123.6  \\ 
5543.944 &  FE1 &  144.4 &  124.4 &  105.8 &  154.9  \\ 
5546.514 &  FE1 &  131.1 &  108.9 &   84.2 &  168.4  \\ 
5560.207 &  FE1 &  120.9 &  100.3 &   80.9 &   --    \\ 
5577.013 &  FE1 &   --   &   --   &   --   &   34.7 \\ 
5587.573 &  FE1 &   --   &   --   &   --   &   --   \\
5635.824 &  FE1 &   --   &   --   &   --   &  117.8 \\ 
5636.705 &  FE1 &   69.9 &   62.4 &   --   &  116.7  \\ 
5638.262 &  FE1 &  180.4 &  155.6 &  140.3 &  191.6  \\
5641.436 &  FE1 &  151.9 &  132.4 &  108.7 &  182.8  \\ 
5646.697 &  FE1 &   --   &   --   &   --   &   58.2  \\ 
5650.019 &  FE1 &   80.5 &   72.2 &   --   &   --    \\
5652.319 &  FE1 &   65.1 &   58.2 &   35.8 &   97.5  \\ 
5661.348 &  FE1 &   72.4 &   62.7 &   33.9 &  128.2  \\ 
5680.240 &  FE1 &   --   &   --   &   --   &   86.8 \\
5701.557 &  FE1 &  227.9 &  192.4 &  160.9 &  --    \\ 
5705.473 &  FE1 &   95.4 &   82.5 &   58.5 &  121.0 \\  
5731.761 &  FE1 &  130.5 &  116.3 &   92.1 &  154.5 \\  
5738.240 &  FE1 &   --   &   --   &   --   &   88.4 \\ 
5778.463 &  FE1 &   87.8 &   74.8 &   38.4 &  160.5 \\ 
5784.666 &  FE1 &   --   &   73.9 &   45.4 &  140.1 \\
5811.916 &  FE1 &   30.9 &   35.4 &   --   &   68.4 \\ 
5814.805 &  FE1 &   58.2 &   56.5 &   35.5 &   96.8 \\ 
5835.098 &  FE1 &   --   &   --   &   --   &   95.7 \\
6012.212 &  FE1 &   --   &   --   &   --   &  168.8 \\ 
6079.014 &  FE1 &  106.4 &   93.1 &   64.8 &  121.3 \\ 
6084.10  &  FE2 &  136.6 &  107.6 &  133.7 &   78.6 \\
6093.666 &  FE1 &   74.6 &   65.2 &   --   &   --   \\  
6098.250 &  FE1 &   --   &   --   &   --   &   98.6 \\ 
6113.33  &  FE2 &  102.6 &   81.9 &   89.4 &   --   \\
6120.249 &  FE1 &   --   &   40.7 &   --   &  183.2 \\ 
6129.70  &  FE2 &   --   &   --   &   64.0 &   --   \\ 
6136.615 &  FE1 &   --   &   --   &   --   &   --   \\
6137.002 &  FE1 &   --   &   --   &   --   &   --   \\ 
6137.702 &  FE1 &  349.2 &  290.0 &  252.9 &  457.4 \\ 
6141.03  &  FE2 &   --   &   --   &   --   &   --   \\  
6149.25  &  FE2 &  172.1 &  137.2 &  171.2 &   90.9 \\ 
6151.616 &  FE1 &  161.8 &   --   &   94.0 &  243.0 \\ 
6157.733 &  FE1 &  176.6 &  153.8 &  120.0 &  230.2 \\  
6159.382 &  FE1 &   39.6 &   38.0 &   22.2 &   91.0 \\  
6165.363 &  FE1 &  111.6 &  101.0 &   70.3 &  150.3 \\  
6170.500 &  FE1 &  188.0 &  158.2 &  130.0 &  --   \\ 
6173.341 &  FE1 &  216.6 &  176.3 &  142.7 &  297.3 \\  
6179.39  &  FE2 &   --   &   --   &   --   &  --    \\ 
6187.987 &  FE1 &  115.1 &   --   &   76.1 &  153.0 \\ 
6191.558 &  FE1 &   --   &  265.4 &   --   &   --  \\
6213.428 &  FE1 &  239.5 &  196.6 &  165.2 &   --  \\ 
6226.730 &  FE1 &   83.5 &   72.7 &   55.3 &  132.4 \\ 
6238.39  &  FE2 &  198.8 &  154.7 &  191.5 &   --   \\
6239.95  &  FE2 &   --   &   --   &  103.0 &   --   \\ 
6240.645 &  FE1 &  171.7 &  138.8 &   99.8 &  239.5  \\ 
6247.55  &  FE2 &  234.8 &  185.1 &  240.7 &  103.7 \\
6248.90  &  FE2 &   --   &   --   &   36.8 &   --   \\  
6271.283 &  FE1 &   --   &   --   &   --   &  156.8 \\ 
6290.974 &  FE1 &  143.4 &   --   &   --   &  174.0 \\
6293.933 &  FE1 &   --   &   --   &   --   &   --   \\ 
6297.799 &  FE1 &  267.5 &  189.3 &  150.1 &  320.9 \\  
6301.508 &  FE1 &  248.5 &  213.8 &  209.8 &  264.3  \\ 
6302.499 &  FE1 &  204.6 &  182.1 &  149.3 &  228.4  \\ 
6305.314 &  FE2 &   --   &   --   &   --   &   --   \\ 
6307.854 &  FE1 &   --   &   --   &   --   &   68.6  \\ 
6315.314 &  FE1 &   --   &  150.8 &   --   &  206.4 \\ 
6315.814 &  FE1 &  112.1 &  109.3 &   77.5 &  150.6 \\ 
6322.694 &  FE1 &  224.6 &  180.6 &  147.2 &  277.3  \\ 
6331.95  &  FE2 &   --   &   --   &   53.6 &   --   \\ 
6336.823 &  FE1 &   --   &  197.6 &  184.6 &  270.0  \\ 
6369.46  &  FE2 &  119.5 &   97.9 &  117.5 &   58.5 \\
6380.750 &  FE1 &  232.0 &  111.4 &   82.1 &   --   \\ 
6383.72  &  FE2 &   --   &   --   &   49.7 &   --   \\ 
6385.45  &  FE2 &   --   &   --   &   --   &   --  \\
6385.726 &  FE1 &   --   &   --   &   --   &   58.7 \\ 
6392.538 &  FE1 &   74.3 &   66.6 &   33.3 &   --   \\ 
6393.602 &  FE1 &   --   &   --   &  246.8 &   --   \\
6400.000 &  FE1 &   --   &   --   &   --   &   --   \\  
6411.658 &  FE1 &  262.4 &  225.2 &  214.3 &  271.6 \\ 
6416.92  &  FE2 &  165.5 &  134.1 &  175.3 &   93.8 \\ 
6419.956 &  FE1 &  203.6 &  164.4 &  144.8 &  208.4  \\ 
6421.360 &  FE1 &  344.2 &  299.2 &  236.0 &  387.9  \\ 
6430.856 &  FE1 &  362.2 &  268.5 &  238.3 &  499.0  \\ 
6432.68  &  FE2 &  186.6 &  154.4 &  189.4 &   97.5 \\  
6442.94  &  FE2 &   --   &   --   &   30.2 &   --   \\ 
6446.40  &  FE2 &   --   &   --   &   --   &   --    \\ 
6456.39  &  FE2 &  268.9 &  214.9 &  277.8 &  108.4  \\ 
\hline
\end{longtable}
%



\begin{thebibliography}{}

\bibitem[2001]{AP01} Allende Prieto, C., Lambert, D.L., Asplund, M., 2001, ApJ, 556, L63.
\bibitem[1998]{A98} Alonso, A., Arribas, S., Martinez-Roger, C., 1998, A\&AS,
131, 209.
\bibitem[1999]{Al99} Alonso, A., Arribas, S., Mart\'{\i}nez-Roger, C., 1999, A\&AS, 140, 261.
\bibitem[1996]{A96} Andrievsky, S.M., Kovtyukh, V.V., Usenko, I.A., 1996, A\&A, 305, 551.
\bibitem[2002]{A02} Andrievsky, S.M., Kovtyukh, V.V., Luck, R.E., L\'epine, J.R.D., Maciel, W.J., Beletsky, Yu. V., 2002, A\&A, 392, 491.
\bibitem[2004]{A04} Andrievsky, S.M., Luck, R.E., Martin, P., L\'epine, J.R.D., 2004, A\&A, 413, 159.
\bibitem[1990]{AFP90} Arellano Ferro, A., Parrao, L., 1990, A\&A, 239, 205.
\bibitem[1983]{BL83} Boyarchuk, A.A., Lyubimkov, L.S., 1983, Izv. Krym. Astrofiz. Obs., 66, 130.
\bibitem[1985]{B85} Barbuy, B., 1985, A\&A, 151, 189.
\bibitem[1996]{B96} Barbuy, B., De Medeiros, J.R, Maeder, A., 1996, A\&A, 305, 911.
\bibitem[2003]{B03} Barbuy, B., Perrin, M.N., Katz, D., Coelho, P., Cayrel, R., Spite, M., Van't Veer-Menneret, C., 2003, A\&A, 404, 661.
\bibitem[2000]{C00} Castilho, B.V., Pasquini, L., Allen, D.M., Barbuy, B., Molaro, P., 2000, A\&A, 361, 92.
\bibitem[1998]{Chen98} Chen, B., Vergely, J.L., Valette, B., Carraro, G., 1998, 
A\&A, 336, 137.
\bibitem[1999]{C99} Cram, L., 1999, Trans. IAU XXIIIB, 1999, 141, Editor: Andersen, J..
\bibitem[Cutri et al. 2003]{2mass} Cutri, R.M. et al., 2003,
{\it The 2Mass All-Sky Catalogue of Point Sources}, University of Massachusetts 
and Infrared Processing and Analysis Center (IPAC/California Institute of Technology)
\bibitem[2002]{M02} De Medeiros, J.R., Udry, S., Burki, G., Mayor, M. 2002
, A\&A, 395, 97.
\bibitem[1995]{EC95} El Eid, M.F., Champagne, A.E., 1995, ApJ, 451, 298.
\bibitem[ESA 1997]{hip} ESA, 1997, {\it The Hipparcos and Tycho catalogues}, ESA SP-1200.
\bibitem[1996]{F96} Fliegner, J., Langer, N., Venn, K., 1996, A\&AL, 308, 13.
\bibitem[1981]{F81} Foy, R., 1981, A\&A, 103, 135.
\bibitem[1992]{GL92} Gies, D., Lambert, D., 1992, ApJ, 387, 673.
\bibitem[1989]{G89} Gilroy, K., 1989, ApJ, 347, 835.
\bibitem[1998]{G98} Grevesse, N., Sauval, A.J., 1998, Space Sci. Rev., 85, 161.
\bibitem[1975]{G75} Gustafsson, B., Bell, R., Eriksson, K., Nordlund, A., 1975, A\&A, 43, 407.
\bibitem[1997]{HK97} Hakkila, J., Myers, J.M., Stidham, B.J., 1997, AJ, 114, 2043.
\bibitem[1995]{H95} Hill, V., Andrievsky, S., Spite, M., 1995, A\&A, 293, 347.
\bibitem[2000]{H00} Houdashelt, M.L., Bell, R.A., Sweigart, A.V., 2000, AJ, 
119, 1448.
\bibitem[2000]{JM00} Jerzykiewicz, M., Molenda-Zacowicz, J., 2000, AcA, 50, 369.
\bibitem[2000]{K00} Kaufer, A., Stahl, O., Tubbesing, S., Norregaard, P., Avila, G., Francois, P., Pasquini, L.,  Pizzella, A., 2000, {\it Performance report on FEROS, the new fiber-linked echelle spectrograph at the ESO 1.52-m telescope}, Proc. SPIE Vol. 4008, Optical and IR Telescope Instrumentation and Detectors, Masanori Iye; Alan F. Moorwood; Eds., 459.
\bibitem[1996]{K96} Kovtyukh, V.V., Andrievsky, S.M., Usenko, I.A., Klochkova, V.G., 1996, A\&A, 316, 155.
\bibitem[1999]{KA99} Kovtyukh, V.V., Andrievsky, S.M., 1999, A\&A, 351, 597.
\bibitem[1994]{K94} Kur\'ucz, R.L., 1994, CD-ROM 19.
\bibitem[1996]{L96} Lennon, D., Dufton, P., Mazzali, P., Pasian, F., Marconi, G., 1996, A\&A, 314, 243.
\bibitem[2003]{L03} Luck, R.E., Gieren, W.P., Andrievsky, S.M., Kovtyukh, V.V., Fouqu\'e, P., Pont, F., Kienzle, F., 2003, A\&A, 401, 939.
\bibitem[1977]{L77} Luck, R.E., 1977, ApJ, 218, 752.
\bibitem[1982]{L82} Luck, R.E., 1982, ApJ, 256, 177.
\bibitem[1980]{LB80} Luck, R.E., Bond, H., 1980, ApJ, 241, 218.
\bibitem[1985]{LL85} Luck, R.E., Lambert, D.L., 1985, ApJ, 298, 782.
\bibitem[1985]{LL92} Luck, R.E., Lambert, D.L., 1992, ApJS, 79, 303.
\bibitem[1998]{L98} Luck, R.E., Moffet, T., Barbes III, T., Gieren, W., 1998, AJ, 115, 605.
\bibitem[1998]{Ly98} Lyubimkov, L., 1998, Astr. Rep., 387, 673.
\bibitem[1983]{LB83} Lyubimkov, L., Boyarchuk, A.A., 1983, Astrofizika, 19, 683.
\bibitem[1997]{M97} Maeder, A., 1997, A\&A, 321, 134.
\bibitem[1998]{MZ98} Maeder, A., Zahn, J.P., 1998, A\&A, 334, 1000.
\bibitem[2000]{MaM00} Maeder, A., Meynet, G., 2000, ARA\&A, 38, 143.
\bibitem[2002]{NIST} Martin, W.C., Fuhr, J.R., Kelleher, D.E., et al. 2002, NIST Atomic Database (version 2.0), http://physics.nist.gov/asd. National Institute of Standards and Technology, Gaithersburg, MD.
\bibitem[2004]{M04} Mathis, S., Palacios, A., Zahn, J.P., 2004, A\&A, 425, 243.
\bibitem[2004]{MZ04} Mathis, S., Zahn, J.P., 2004, A\&A, 425, 229.
\bibitem[1991]{MW91} McWilliam, A. 1991, AJ, 101, 1065.
\bibitem[2005]{MB05} Mel\'endez, J., Barbuy, B., 2005, in preparation.
\bibitem[2000]{MeM00} Meynet, G., Maeder, A., 2000, A\&A, 361, 101.
\bibitem[1992]{M92} Milone, A., Barbuy, B., Spite, M., Spite, F., 1992, A\&A, 261, 551.
\bibitem[1992]{P92} Plez, B. and Brett, J.M. and Nordlund, A., 1992, A\&A, 256, 551.
\bibitem[1967]{P67} Praderie, F., 1967, Ann. Astrophys., 30, 31.
\bibitem[1985]{RL85} Rieke, G.H., Lebofsky, M.J., 1985, ApJ, 288, 618.
\bibitem[2002]{Ro02} Royer, F., Gerbaldi, M., Faraggiana, R., G\'omez, A., 2002, A\&A, 381, 105.
\bibitem[1986]{S86} Sasselov, D.D., 1986, PASP, 98, 561.
\bibitem[1992]{Sc92} Schaller, G., Schaerer, D., Meynet, G., Maeder, M., 1992, A\&AS, 96, 269.
\bibitem[1989]{S89} Spite, M., Barbuy, B., Spite, F., 1989, A\&A, 222, 35.
\bibitem[1998]{TT98} Takeda, Y., Takada-Hidai, M., 1998, PASJ, 50, 629.
\bibitem[2001a]{U01a} Usenko, I.A., Kovtyukh, V.V., Klochkova, V.G., 2001a, A\&A, 377, 156.
\bibitem[b]{U01b} Usenko, I.A., Kovtyukh, V.V., Klochkova, V.G., Panchuk, V.E., Yermakov, S.V., 2001b, A\&A, 367, 831.
\bibitem[1999]{VB99} van Belle, G.T., Lane, B.F., Thompson, R.R. et al., 1999, AJ,
117, 521.
\bibitem[1995a]{V95a} Venn, K., 1995a, ApJ, 449, 839.
\bibitem[b]{V95b} Venn, K., 1995b, ApJS, 99, 659.
\bibitem[1999]{V99} Venn, K., 1999, ApJ, 518, 405.
\bibitem[2002]{Venn02} Venn, K., Brooks, A.M., Lambert, D.L., Lemke, M., Langer, N., Lennon, D.J., Keenan, F.P., 2002, ApJ, 565, 571.
\bibitem[2004]{XL04} Xu, H.Y., Li, Y., 2004, A\&A, 418, 213.

\end{thebibliography}
\end{document}